\documentclass[
  reprint,
  10pt,
  aps,
  prx,
  twocolumn,
  superscriptaddress,
  amsmath,
  amssymb,
  floatfix,
  showkeys,
]{revtex4-2}

\usepackage[dvipsnames]{xcolor}

\usepackage{graphicx}
\usepackage{dcolumn}
\usepackage{bm}
\usepackage{subfiles}
\usepackage{orcidlink}
\hypersetup{
  colorlinks=true,
  allcolors=blue
}

\definecolor{mutedblue}{RGB}{40,60,110}
\definecolor{deepred}{RGB}{150,40,40}
\definecolor{accentblue}{HTML}{069AF3}

\begin{document}

\title{\texorpdfstring{
A Minimal Model of Representation Collapse:\\
Frustration, Stop-Gradient, and Dynamics
}{A Minimal Model of Representation Collapse: Frustration, Stop-Gradient, and Dynamics}}

\author{Louie Hong Yao\,\orcidlink{0000-0001-6910-2951}}
\email{lhyao731@gmail.com}
\thanks{Equal contribution}
\affiliation{Center for Soft Matter and Biological Physics, Virginia Tech, Blacksburg, VA 24061, USA}

\author{Yuhao Li\,\orcidlink{0009-0000-6592-5475}}
\thanks{Equal contribution}
\affiliation{Department of Computer Science and Engineering, The Chinese University of Hong Kong, Hong Kong, China}

\author{Shengchao Liu\,\orcidlink{0000-0003-2030-2367}\,}
\email{scliu@cuhk.edu.hk}
\affiliation{Department of Computer Science and Engineering, The Chinese University of Hong Kong, Hong Kong, China}

\date{\today}

\begin{abstract}
Self-supervised representation learning is central to modern machine learning because it extracts structured latent features from unlabeled data and enables robust transfer across tasks and domains.
However, it can suffer from representation collapse, a widely observed failure mode in which embeddings lose discriminative structure and distinct inputs become indistinguishable.
To understand the mechanisms that drive collapse and the ingredients that prevent it, we introduce a minimal embedding-only model whose gradient-flow dynamics and fixed points can be analyzed in closed form, using a classification-representation setting as a concrete playground where collapse is directly quantified through the contraction of label-embedding geometry. 
We illustrate that the model does not collapse when the data are perfectly classifiable, while a small fraction of frustrated samples that cannot be classified consistently induces collapse through an additional slow time scale that follows the early performance gain.
Within the same framework, we examine collapse prevention by adding a shared projection head and applying stop-gradient at the level of the training dynamics.
We analyze the resulting fixed points and develop a dynamical mean-field style self-consistency description, showing that stop-gradient enables non-collapsed solutions and stabilizes finite class separation under frustration.
We further verify empirically that the same qualitative dynamics and collapse-prevention effects appear in a linear teacher-student model, indicating that the minimal theory captures features that persist beyond the pure embedding setting.
\end{abstract}

\maketitle


\section{Introduction}

Recent advances in artificial intelligence (AI) have delivered striking performance across a wide range of tasks, from perception~\cite{oquab2024dinov,simeoni2025dinov3,Rombach2022high} and language understanding \cite{liu2025deepseek,guo2025deepseek, yang2025qwen3, wei2025deepseek, comanici2025gemini} to scientific and engineering applications \cite{jumper2021highly, google2025quantum, novikov2025alphaevolve, liu2025multi, avsec2025alphagenome}.
Much of this progress is driven by large-scale data, large models, and effective training recipes, which together yield strong generalization and rapid adaptation to new domains.
However, the same ingredients introduce an enormous number of degrees of freedom and a training process that is highly coupled, nonlinear, and often messy in its details, so a clear theoretical picture remains limited.
At the same time, physics has long developed tools to analyze complex many-body systems, where qualitative behavior emerges from collective effects rather than microscopic details~\cite{anderson1972more}.
This perspective naturally suggests that modern learning systems may also admit robust effective theories.

In line with this idea, physics has interacted with modern AI in two complementary directions.
On one side, physicists have brought analytical tools together with physical intuition to understand learning systems.
Examples include nonequilibrium-dynamics viewpoints on optimization~\cite{feng2021inverse, yang2026transient}, mean-field and dynamical mean-field approaches for high-dimensional networks and their representation dynamics~\cite{sompolinsky1988chaos, cowsik2025geometric, epping2024graph}, renormalization-group and symmetry viewpoints on generative model behavior~\cite{kamb2024analytic, sheshmani2025renormalization}, and spin-glass perspectives on emergent computation in large models~\cite{hou2025focused, li2025spin, barney2024neural}.
On the other side, ideas from physics have shaped both model classes and training principles in modern AI.
For instance, thermodynamic and dynamical formulations motivate diffusion and flow-based generative models~\cite{sohl2015deep, wang2025equilibrium, du2025flow}.
Related perspectives draw on phase transitions and spontaneous symmetry breaking to organize generative dynamics across scales~\cite{yaguchi2025the, raya2023spontaneous}, and on synchronization phenomena to inspire coupled-oscillator style architectures~\cite{song2025kuramoto, miyato2025artificial}.

\begin{figure*}[t]
    \centering
    \includegraphics[width=0.98\linewidth]{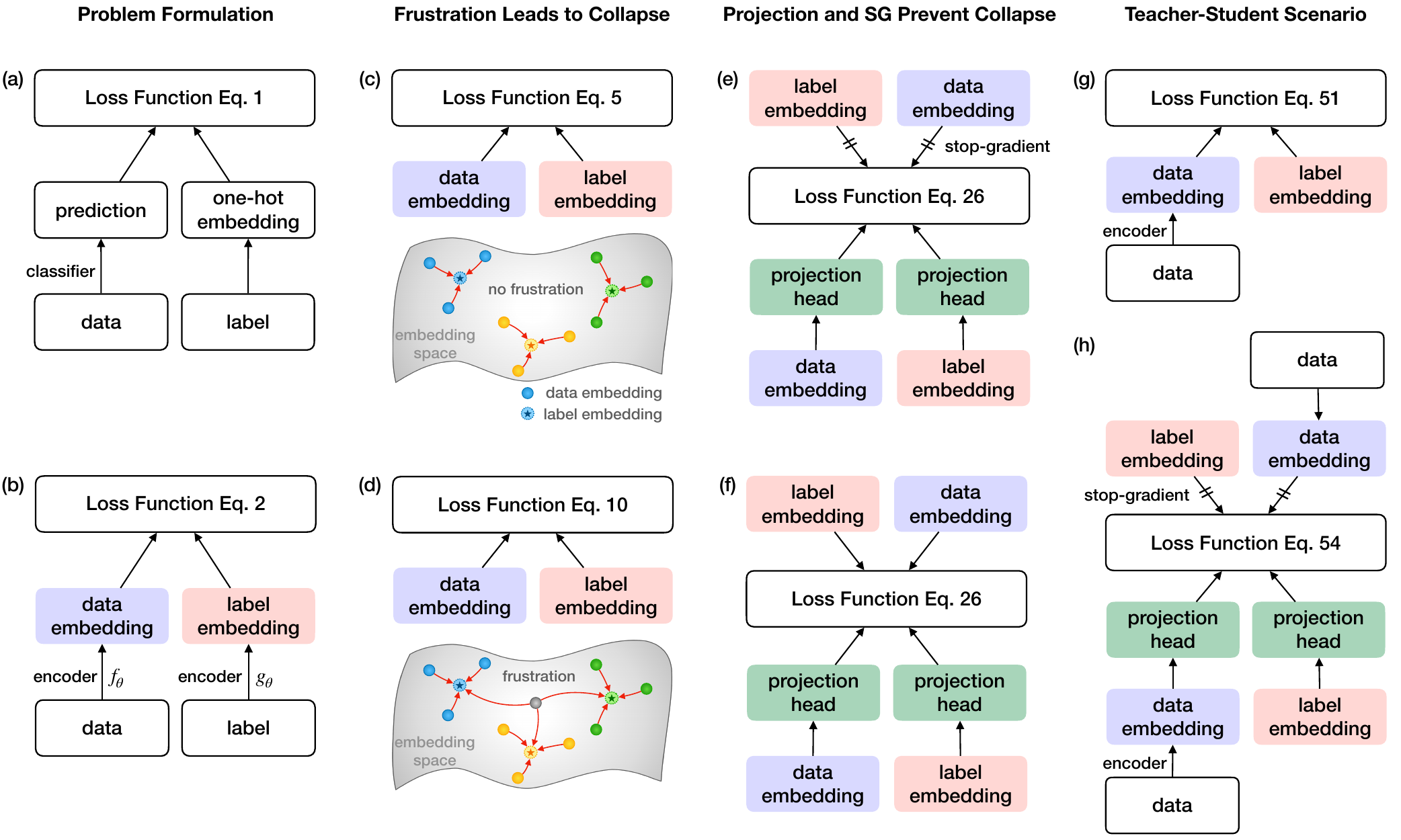}
    \caption{
    Overview of model collapse in representation learning.
    (a) Standard contrastive-based classification pipeline with fixed one-hot labels, where only data representations are learned.
    (b) Generative-based representation-learning pipeline used in this work, with learnable data and label embeddings.
    (c) Embedding model under Eq.~\eqref{eq:minimal-mse} without frustration, where class-wise dynamics remain decoupled.
    (d) Same model with frustration induced by shared data embeddings, leading to competing alignment constraints across classes.
    (e) Architecture with projection heads and stop-gradient [Eq.~\eqref{eq:simsiam_minimal}], which modifies the effective gradient flow between branches.
    (f) Same architecture without stop-gradient, restoring symmetric coupling between embeddings.
    (g) Teacher–student pipeline where data embeddings are generated by a parametrized encoder.
    (h) Teacher–student model with a learned encoder, projection head, and stop-gradient, extending the mechanism beyond the embedding-only setting.
    }
    \label{fig:overview}
\end{figure*}

Despite these advances, many concrete phenomena in learning systems still lack a clean dynamical account that isolates which ingredients control qualitative outcomes.
A prominent example is representation collapse in self-supervised representation learning~\cite{balestriero2024the, shwartz2024compress,lecun2021darkmatter}, which aims to assign each data sample a vector representation whose embedding geometry captures meaningful relations and supports transfer to downstream tasks.
In this setting, collapse refers to a degeneration of the embedding geometry, where representations lose discriminative structure and many distinct inputs are mapped to nearly the same point.
A range of strategies have been proposed to prevent this outcome.
Explicit approaches enforce non-collapse by construction, for example through contrastive objectives with negative pairs~\cite{chopra2005learning, oord2018representation, chen2020simclr} or through explicit covariance regularization~\cite{Assran2023jepa}.
More intriguingly, implicit approaches such as BYOL~\cite{grill2020byol} and SimSiam~\cite{Chen2021simsiam} do not impose an explicit repulsive force, yet still prevent collapse through architectural or dynamical asymmetries. This raises the basic questions of when and why collapse arises in the first place and how these methods mitigate it through their effects on the dynamics.

Previous studies have sought to explain why implicit methods avoid collapse mainly in two ways, either through extensive empirical simulations~\cite{wang2022importance, wang2021towards} or through simplified toy networks with specific architectures~\cite{tian2021understanding, zhang2022how}.
When theory is pursued, it typically proceeds by simplifying the network, for instance to two-layer models, to obtain a tractable description of the training dynamics.
While these analyses are valuable, they often remain closely tied to microscopic architectural details, making it difficult to distill the dynamics into a simple effective theory.
One reason is that even in such simplified settings the analysis typically starts from a detailed parameter-level, microscopic or ultraviolet (UV) description and then attempts to derive the emergent infrared (IR) effective theory from it.

In this work, we take a complementary route and formulate a minimal theory directly at the infrared level.
Rather than starting from ultraviolet variables such as specific weights and architectural constraints, we treat the embeddings themselves as the effective degrees of freedom and focus on the collective dynamics they obey.
We study these dynamics within a classification playground, where sample embeddings are trained to match labels. Unlike mainstream approaches, where representations are trained to match labels in a categorical space, we instead consider alignment in a learnable label embedding space, where collapse is quantified directly through the contraction of the label-embedding geometry.
An overview is illustrated in Fig.~\ref{fig:overview}.

This representation-level formulation admits closed-form dynamics and fixed-point analysis, and it shows that when all samples are perfectly classifiable, the learned structure can persist without collapse.
Collapse is instead induced by frustration, defined as the fraction of samples that cannot be classified consistently. 

Such frustration is not a modeling artifact, but a generic feature of realistic learning problems, where it may arise from imperfect data, label noise, or limited model expressivity. At the representation level, however, these different sources share the same effective consequence, with some embeddings failing to align consistently with their class labels.
When the fraction of such frustrated samples is small, it introduces a separate slow time scale, resulting in rapid early improvement followed by late-time degradation.
This slow time scale is governed by the frustration fraction: as the fraction increases, it approaches the fitting time scale, and the separation between the two stages gradually vanishes.

Within the same framework, we also study how techniques such as adding a shared projection head and applying stop-gradient mitigate collapse.
We show that stop-gradient enables non-collapsed fixed points by opening non-collapsed directions in the representation space.
From the Dynamical Mean Field Theory (DMFT) self-consistency equations, we further find that this non-collapsed eigensubspace emerges from the interaction between the projection head and frustration through the structure of the propagator.

Moreover, to test whether these mechanisms persist once one restores a learned input-to-representation map, we validate the results in a linear teacher-student model, which reintroduces a parametrized mapping from inputs to embeddings.
The same qualitative time-scale separation and stop-gradient stabilization persist beyond the pure embedding setting.
Taken together, our minimal model identifies frustration as the driving ingredient behind collapse and clarifies how stop-gradient avoids it by enabling non-collapsed fixed points through additional non-collapsing directions in representation space.

The paper is organized as follows.
We begin in Sec.~\ref{sec:background} with the motivating empirical observations from classification-representation training and define the geometric quantities we use to diagnose collapse through the behavior of label embeddings.
Sec.~\ref{sec:minimal model collapse} and~\ref{sec:simsiam_minimal} then develop the minimal model and use it to study both sides of the story, how frustration produces collapse through the fixed points and time scales and how a shared projection head with stop-gradient opens up non-collapsed fixed points and alters the dynamics.
To connect back to a more standard learning setup, Sec.~\ref{sec:teacher-student} revisits the same phenomena in a linear teacher--student framework.
We finish in Sec.~\ref{sec:conclusion} with a brief discussion of implications, limitations, and future directions.


\section{Mode Collapse in Representation Learning}
\label{sec:background}
In this paper, we focus on the classification problem and later analyze it in a simplified setting, where the collapse phenomenon becomes easier to detect.
Fig.~\ref{fig:overview}(a) shows a standard classification problem, which is characterized as a supervised problem. We are given data points $\{\mathbf{x}_{\alpha,i}\}$, where $\alpha = 1,2,\ldots,N$ labels the class to which each data point belongs. A parametrized function $f_\theta(\cdot)$ is then learned by optimizing the parameters $\theta$ over a loss function $\mathcal{L}$, 
such as the typical contrastive loss, cross-entropy:
\begin{equation}
    \mathcal{L}_\text{class} = \sum_{\alpha,i} \text{CrossEntropy}\Big(f_\theta(\mathbf{x}_{\alpha i}), y_\alpha\Big),
\end{equation}
where $y_\alpha$ is the corresponding one-hot label of class $\alpha$. In this setting, the labels are fixed as one-hot vectors in the space $\mathbb{R}^N$, with $N$ denoting the number of classes. The optimization over $\theta$ is to find a specific function $f_\theta(\cdot)$ that makes $f_\theta(x_{\alpha i})$ as close to $y_\alpha$ as possible. Since the $y_\alpha$ are fixed, there is no issue of mode collapse.

Here we depart from the traditional contrastive-based classification setup and consider a generative-based representation-learning variant, as shown in Fig.~\ref{fig:overview}(b).
In this case, we further employ an encoding function for the class labels $g_\theta(\cdot)$, and the loss function becomes a functional of both $f_\theta$ and $g_\theta$. We write $f_\theta$ and $g_\theta$ for the data and label encoders respectively, and let $\theta = (\theta_f,\theta_g)$ denote the collection of all trainable parameters. In particular, $f$ and $g$ do not share model architecture or weights; $\theta$ is only a shorthand for the joint parameter space. 

To train the model, we consider two types of loss.
The first is the mean-square-error (MSE) loss, corresponding to the matching in the Euclidean distance.
Another common choice is the negative cosine loss, which instead measures distance on the unit sphere. Given the MSE loss, the functions $f_\theta$ and $g_\theta$ are optimized over
\begin{equation}
    \mathcal{L}_\text{MSE} = \sum_{\alpha, i} \Big(f_\theta(x_{\alpha i}) - g_\theta(y_\alpha)\Big)^2.
\label{eq:mse loss}
\end{equation}
This setting naturally appears in multiscale learning in the physical sciences \cite{maggioni2016multiscale, kalimuthu2025loglofno, liu2025multi, parsan2025towards}, where, due to scale separation and universality, different microscopic settings correspond to similar mesoscopic and/or macroscopic behaviors. Moreover, when the class labels are taken to be the data points themselves and the $x$ are augmented versions of these data points, it reduces to the multi-view learning setting \cite{lecun2022path} in computer vision and natural language processing.

Given the loss in Eq.~\eqref{eq:mse loss}, we optimize the parameters $\theta$ by (stochastic) gradient descent with learning rate $\eta$. For a mini-batch $\mathcal{B}$, one update step takes the form
\begin{equation}
    \theta \leftarrow \theta - \eta \nabla_\theta \mathcal{L}_\text{MSE}(\mathcal{B}).
\end{equation}
In many settings, one additionally includes weight decay, which is equivalent to augmenting the loss by an $L_2$ regularization term $\frac{\lambda}{2}\|\theta\|_2^2$. We will not discuss that case further here, as our goal is to isolate the intrinsic collapse dynamics without introducing an additional decay time scale.

\subsection{Mode collapse}
Since the label representations $g_\theta(y)$ are not fixed in the representation space, there is a natural global minimum of Eq.~\eqref{eq:mse loss}, achieved when $f_\theta(x) = \mathbf{C} = g_\theta(y)$ for all $x$ and $y$, where $\mathbf{C}$ is a constant vector. In this case, $\mathcal{L}_\text{MSE}=0$, achieving the global minimum.
The optimizer can reach this state by driving every representation toward the same vector $\mathbf{C}$, so that all pairwise differences vanish and the dataset collapses to a single point.
At this collapsed solution, class structure is erased: any classifier built on the representations cannot separate classes, and the accuracy reduces to chance level (e.g., $1/N$ for balanced $N$-way classification).
Besides the complete mode collapse that gives the global minimum, partial mode collapse can also occur in this classification setting, for example, when $g_\theta(\alpha_1) = g_\theta(\alpha_2)$ for $\alpha_1 \neq \alpha_2$. In this case, the distinction between class $\alpha_1$ and class $\alpha_2$ is lost, and the classification accuracy will suddenly drop by a step of order $1/N$.

\begin{figure}[ht]
    \centering
    \includegraphics[width=\linewidth]{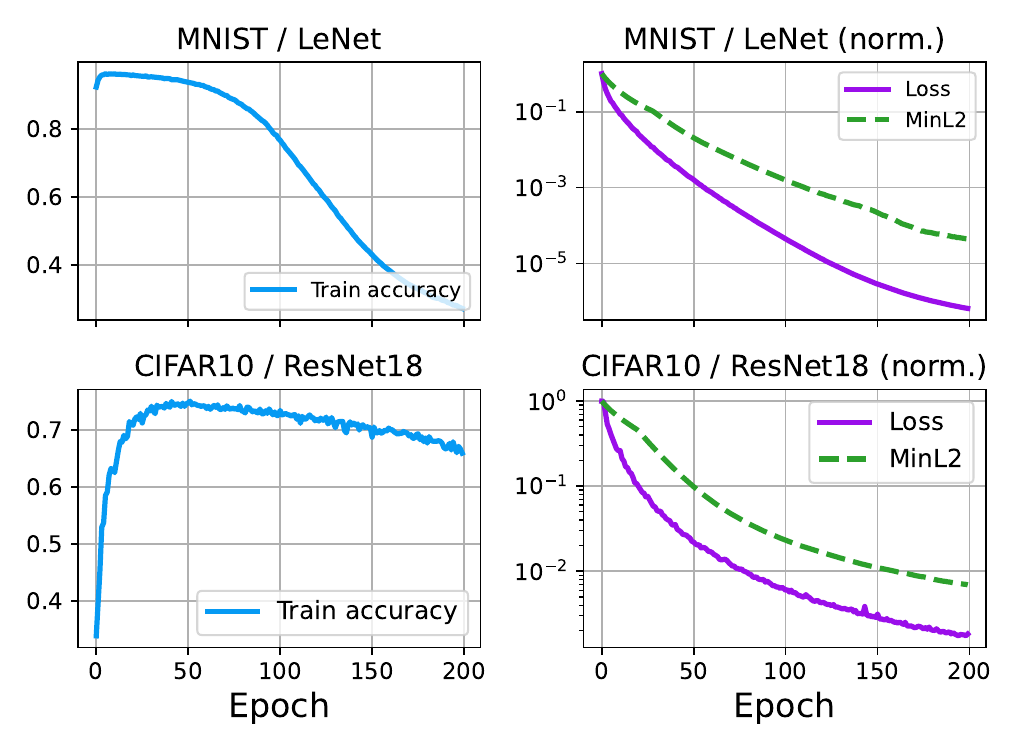}
   \caption{Training dynamics of the classification–representation model on MNIST and CIFAR-10.
    In both experiments, we use an embedding dimension of 10, batch size 128, and stochastic gradient descent without weight decay or other regularization; the learning rate is $0.05$ for MNIST and $0.01$ for CIFAR-10.
    Left: training accuracy as a function of epochs for MNIST with LeNet and CIFAR-10 with ResNet-18.
    Right: corresponding training loss and minimal $L_2$ distance between label embeddings (\textit{MinL2}), both normalized by their values at epoch zero and shown on a logarithmic scale, highlighting a fast accuracy-improvement regime followed by a slower collapse-dominated regime.
    }
    \label{fig:mnist_cifar_cl}
\end{figure}

In Fig.~\ref{fig:mnist_cifar_cl}, we show the results on MNIST dataset with LeNet model~\cite{lecun2002gradient} and on CIFAR-10 dataset~\cite{krizhevsky2009learning} with ResNet model~\cite{he2016deep}. Since the collapse behavior is most apparent on the training set, we focus on the training dynamics only. The training-set accuracy is shown in the left panel: for both MNIST and CIFAR-10, the accuracy increases rapidly in the initial period and then starts to decay. This is generally in contrast to the usual intuition that, with a proper learning rate, continued training on the training set should keep improving performance on the training set and eventually overfit it, especially when the model is overparameterized.

To better understand this behavior, we also plot the loss and the minimal $L_2$ distance between embeddings of any pair of labels,
\textit{MinL2} $ \equiv \min_{\alpha \neq \beta} \lVert g_\theta(y_\alpha) - g_\theta(y_\beta) \rVert_2$,
in the right panel.
Since \textit{MinL2} characterizes the separation between label embeddings, we use it as a metric to capture mode collapse in the training dynamics.

For easier comparison, we normalize both quantities by their values at epoch zero and show them on a logarithmic $y$-axis. As seen in the plot, the loss exhibits a fast initial decay compared to \textit{MinL2}, corresponding to the rapid growth in accuracy. However, after a crossover period, the decay of the loss gradually becomes parallel to the decay of \textit{MinL2}, indicating that the dynamics are essentially governed by the collapsing behavior. This change of regime suggests the existence of two distinct time scales: one that controls the intended performance at early times and another defined by the collapsing dynamics of the system.

Note that in both cases we optimize the system using stochastic gradient descent without weight decay or other regularization terms. Regularizations generally introduce an additional timescale, which could interfere with the two time scales intrinsic to the system.

\subsection{Mitigation of Collapse}
To mitigate representation collapse, both explicit and implicit approaches have been developed in the community.
Explicit approaches prevent collapse by construction, for example, by introducing contrasted examples through contrastive objectives \cite{chopra2005learning, oord2018representation, chen2020simclr} or by explicitly regularizing the embedding covariance, as in JEPA \cite{Assran2023jepa}.
In contrast, implicit approaches such as BYOL \cite{grill2020byol} and SimSiam \cite{Chen2021simsiam} stabilize non-collapsed representations through architectural or dynamical asymmetries, without relying on negative samples.
In this work, we study the stop-gradient as the key dynamical ingredient underlying this class of implicit approaches.

To adapt the stop-gradient mechanism to the classification--representation setting considered here, we introduce a shared projection network $h_\theta(\cdot)$ that is applied symmetrically to both data and label embeddings.
Specifically, we replace the original mean-squared-error objective in Eq.~\eqref{eq:mse loss} with the projected variant
\begin{equation}
\begin{aligned}
\mathcal{L}_\text{MSE}
= &\frac{1}{2}\sum_{\alpha,i}
\Big(
h_\theta\!\big(f_\theta(x_{\alpha i})\big)
-
\mathrm{sg}\!\left[g_\theta(y_\alpha)\right]
\Big)^2
\\ & + \frac{1}{2}\sum_{\alpha,i}
\Big(
h_\theta\!\big(g_\theta(y_\alpha)\big)
-
\mathrm{sg}\!\left[f_\theta(x_{\alpha i})\right]
\Big)^2 ,
\end{aligned}
\label{eq:proj-sym}
\end{equation}
which symmetrically matches projected data embeddings to label embeddings and vice versa.
The stop-gradient operator $\mathrm{sg}[\cdot]$ blocks gradient flow through the target branch during backpropagation, so that the corresponding partial derivatives do not appear in the parameter updates.
Equivalently, in the gradient descent updates, all terms proportional to $\nabla_\theta  \mathrm{sg}[\cdot]$ vanish identically, while the forward value of the target is retained in the loss.
Importantly, this operation does not modify the loss landscape defined by Eq.~\eqref{eq:proj-sym}, which may be viewed as an energy function.
Instead, it changes the gradient-induced vector field that governs the training dynamics.
As a result, the parameter updates no longer correspond to a steepest-descent flow with respect to this loss.

\begin{figure}[t]
    \centering
    \includegraphics[width=\linewidth]{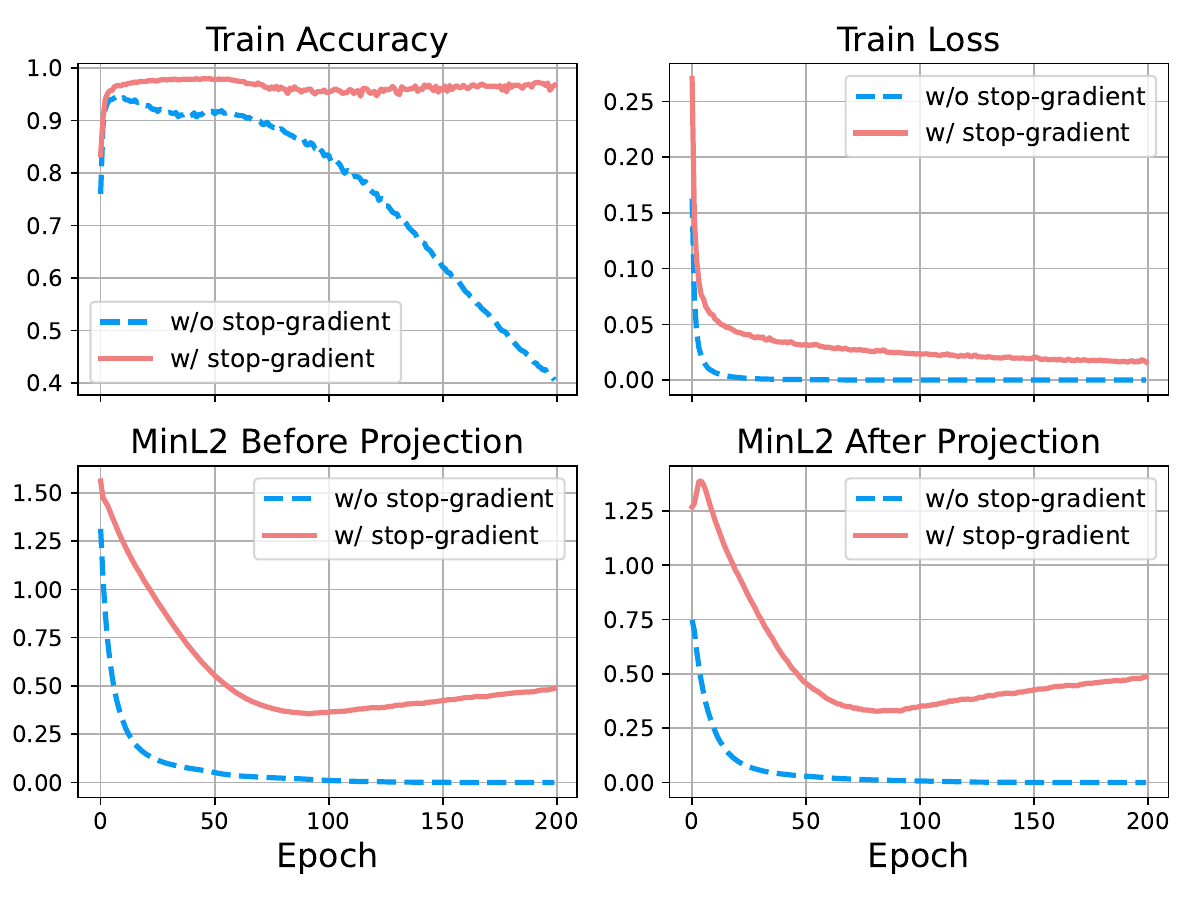}
    \caption{
Effect of stop-gradient on representation collapse.
Training dynamics on MNIST with LeNet under standard optimization and stop-gradient.
Top: training accuracy (left) and loss (right).
Bottom: minimal $L_2$ distance between label embeddings before (left) and after (right) projection.
Stop-gradient preserves early accuracy improvement while suppressing the decay of inter-label distances, stabilizing non-collapsed representations, and preventing the collapse-dominated regime.
    }
    \label{fig:mnist-stopgrad}
\end{figure}

In Fig.~\ref{fig:mnist-stopgrad}, we show the training dynamics of Eq.~\eqref{eq:proj-sym} on MNIST.
The experiment implements the symmetric projected loss with stop-gradient defined in Eq.~\eqref{eq:proj-sym}, where the projection network $h_\theta$ is chosen to be a single linear layer without bias.
All other aspects of the network architecture and optimization settings are kept unchanged.
By tracking both task performance and inter-label separation, the figure shows that even with a minimal linear projection, stop-gradient leaves the early task-learning dynamics intact while suppressing the late-time collapse of label representations.
Without stop-gradient, the training accuracy initially increases but subsequently decreases at late times, even as the training loss continues to decay. Additionally, the minimal inter-label distance \textit{MinL2} decreases monotonically both before and after projection, indicating a progressive collapse of label representations.
When stop-gradient is applied, the training accuracy stabilizes at a high level, while the training loss no longer decreases to the extremely low values observed without stop-gradient.
At the same time, the early-time decay of \textit{MinL2} remains similar with and without stop-gradient, but its late-time behavior changes qualitatively.
Without stop-gradient, \textit{MinL2} continues to decay toward zero; in contrast, with stop-gradient, it saturates at a finite value, indicating the absence of representation collapse.

\noindent
\textbf{Discussion. }
The dynamics shown here correspond to a single training run over a finite time window of 200 epochs.
Because SGD introduces stochasticity, rare collapse events can still occur in single training runs, especially when the dynamics are evolved for long times, even if the overall probability of collapse is low.
Nevertheless, models in the machine learning community are typically trained based on a single run,
\footnote{While ensemble methods such as bagging are common in traditional machine learning, they are rarely applied in neural network training due to computational cost.}
and training is usually terminated well before extremely long-time dynamics are reached.
Within this regime, the results here show that the stop-gradient mechanism can prevent collapse in a meaningful dynamical sense.
In the following section, we introduce a minimal model to provide insight into these dynamics and the mechanisms underlying representation collapse and its mitigation.


\section{Frustration Leads to Representation Collapse}
\label{sec:minimal model collapse}
We adopt a minimal modeling perspective in which the neural network is integrated out entirely, and the dynamics are formulated directly in representation space. This lets us bypass the complexity of network-level training and focus on the effective gradient-flow dynamics of the embeddings. Within this reduced description, collapse is not generic, but is instead driven by frustration: a fraction of samples cannot be classified perfectly by the network and thus induce competing alignment constraints.
Such frustration may arise from imperfect data, label noise, limited model expressivity, or other microscopic causes. In the reduced representation-level description, however, all of these mechanisms enter through the same effect: a subset of embeddings cannot be brought into consistent alignment with their class labels.
We therefore start from the unfrustrated case and then examine how frustration qualitatively changes the dynamics and fixed-point structure.

\subsection{Dynamics without Frustration}
We consider a simplified model of a classification problem with $n$ classes.
Each class is assumed to contain $N$ samples.
We reduce the problem to directly optimizing embeddings for both the data samples and the class labels.
Specifically, we denote by $u_{i\alpha} \in \mathbb{R}^d$ the embedding of the $\alpha$-th sample from class $i$, and by $v_i \in \mathbb{R}^d$ the embedding of the corresponding class label, where $i = 1,2,\ldots,n$ and $\alpha = 1,2,\ldots,N$.
Here $d$ denotes the embedding dimension. 
In this reduced formulation, these embeddings are optimized directly.

As a starting point, we analyze the simplest setting in which there is no coupling between different classes, as shown in Fig.~\ref{fig:overview}(c).
The loss is a mean-squared error between each sample embedding and its corresponding class-label embedding,
\begin{equation}
\ell(\{u\},\{v\})
=\frac{1}{2}\sum_{i=1}^n\sum_{\alpha=1}^N\big(u_{i\alpha}-v_i\big)^2.
\label{eq:minimal-mse}
\end{equation}
Because the loss decomposes as a sum over classes and embedding coordinates, there are no interactions between different classes.
Each class is therefore optimized independently, and embeddings associated with different labels remain completely decoupled.
Moreover, the permutation symmetry among classes implies that all classes relax in the same manner.
The gradient-flow dynamics are derived explicitly in
Appendix~\ref{app:mse-solution}. Here, we focus on the resulting fixed-point
structure.

Within each class, all sample embeddings converge to the corresponding label embedding,
\begin{equation}
u_{i\alpha}=v_i \quad \text{for all }\alpha,
\end{equation}
while the label embedding itself relaxes to a constant determined by the initialization.
Due to the permutation symmetry among samples within a class, the long-time behavior depends only on the within-class sample mean.
In particular, the asymptotic value of the label embedding is given by
\begin{equation}
v_i(\infty)=\frac{N \bar{u}_i(0)+v_i(0)}{N+1},
\label{eq:vinfty}
\end{equation}
where $\bar{u}_i(0)=\frac{1}{N}\sum_{\alpha}u_{i\alpha}(0)$ denotes the initial within-class sample mean.

\begin{figure}[t!]
    \centering
    \includegraphics[width=\linewidth]{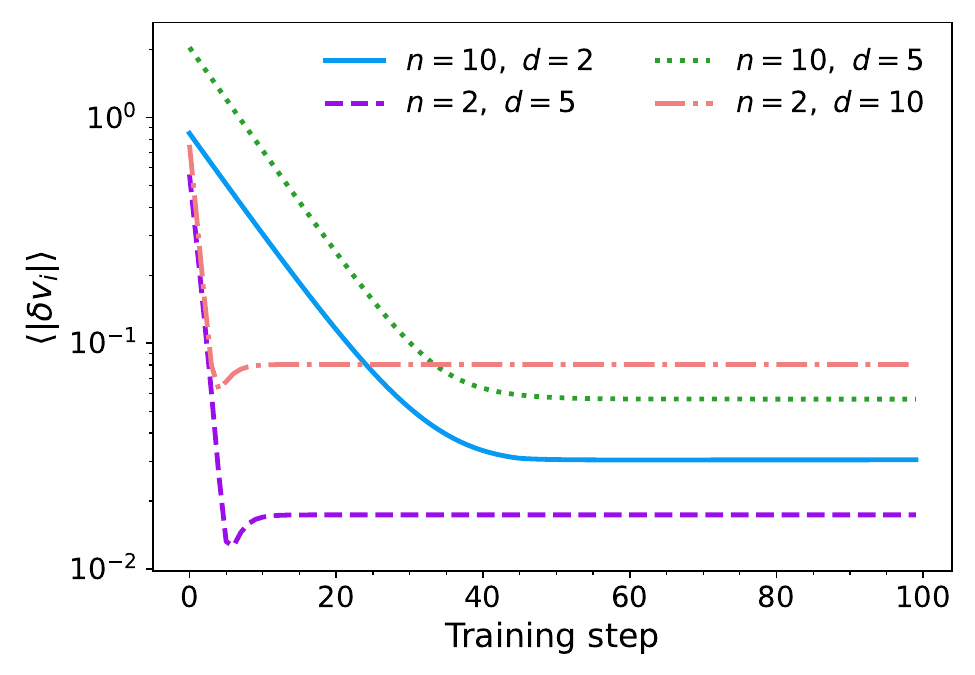}
    \caption{
    Mean inter-class deviation $\langle |\delta v_i| \rangle$ as a function of training step for unfrustrated runs at learning rate $0.5$. The vertical axis is shown on a logarithmic scale. All curves relax rapidly to small but nonzero plateaus, indicating stable class separation rather than complete collapse.
    }
    \label{fig:no_frustration_no_collapse}
\end{figure}

The asymptotic separation can be derived directly as follows. Let
\begin{equation}
\delta v_i = v_i(\infty) - \frac{1}{n}\sum_{j=1}^n v_j(\infty).
\end{equation}
Using Eq.~\eqref{eq:vinfty}, this deviation can be expressed in terms of the initial conditions as:
\begin{equation}
\begin{aligned}
\delta v_i = \frac{N}{N+1} \Big[ \bar{u}_i(0) - \frac{1}{n}\sum_j \bar{u}_j(0) \Big]
\\ + \frac{1}{N+1} \Big[ v_i(0) - \bar{v}(0) \Big],
\end{aligned}
\end{equation}
where $\bar{v}(0) = \frac{1}{n}\sum_j v_j(0)$ denotes the center of the label embedding at initialization.
For finite $N$ the deviation $\delta v_i$ is generically nonzero unless the initial class means and label embeddings are identical across all classes. In particular, the second term in $\delta v_i$ does not vanish under generic Gaussian initialization and therefore prevents exact collapse at finite $N$. 

To confirm this behavior, we directly simulate the dynamics and measure the resulting class separation. As shown in Fig.~\ref{fig:no_frustration_no_collapse} we plot the mean inter class deviation $\langle |\delta v_i| \rangle$ as a function of training time for different numbers of classes $n$ and embedding dimensions $d$. Across all configurations the inter class deviations exhibit a rapid exponential decay at early times and subsequently relax to small but nonzero plateau values, indicating stable class separation throughout training rather than collapse.

Taken together, these observations indicate that in this minimal unfrustrated model, there is no intrinsic mechanism that drives different classes to collapse onto a common representation. Collapse is therefore not a generic outcome of the dynamics.

\subsection{
Dynamics with Frustration and the Emergence of Representation Collapse
}
\label{sec:frustrated dynamics}

So far, we have assumed that embeddings are free to move so as to satisfy all alignment constraints imposed by the loss.
By integrating out the neural network, we effectively allowed the embedding of each sample to evolve independently, without interacting with other samples.
In practice, however, perfect alignment is typically impossible, leaving some samples that cannot be consistently assigned to a single class.

To model this effect at a minimal level, we introduce frustration directly in representation space by allowing a subset of samples to interact with multiple class labels.
For a fraction $1-r$ of the samples, embeddings $u_{i\alpha}$ are paired exclusively with their corresponding class label $v_i$. For the remaining fraction $r$, embeddings $s_\alpha$ are simultaneously paired with all class labels, as shown in Fig.~\ref{fig:overview}(d). 
These shared samples represent data points that cannot be uniquely assigned to a single class 
and therefore impose competing alignment constraints across classes.
The resulting loss function takes the form
\begin{equation}
\ell
= \sum_{i=1}^n \sum_{\alpha=1}^{(1-r)N}
\big(u_{i\alpha}-v_i\big)^2
+ \sum_{i=1}^n \sum_{\alpha=1}^{rN}
\big(s_\alpha-v_i\big)^2 .
\label{eq:frustrated-loss-noproj}
\end{equation}
Unlike the unfrustrated case, this loss no longer decomposes into a sum over independent classes.
Each embedding $s_\alpha$ in the second term is simultaneously constrained to align with multiple label embeddings $v_i$, generating competing forces that cannot, in general, be minimized simultaneously.

We consider full-batch gradient descent with ``raw'' learning rate $\tilde{\gamma}$.
In practice, training proceeds via discrete updates and the loss is averaged
over all samples.
At fixed $N$, the total number of samples scales linearly with the number of
classes $n$, introducing an overall normalization that rescales the gradient
flow by a factor $1/n$.
When analyzing the dynamics, we work in a continuous-time description, in which
both discretization effects and the normalization induced by loss averaging are
absorbed into the definition of an effective learning rate $\gamma$.
Differentiating the loss in Eq.~\eqref{eq:frustrated-loss-noproj} then yields the
equations of motion
\begin{equation}
\begin{aligned}
\dot{u}_{i\alpha}
&= -\gamma (u_{i\alpha}-v_i), \\
\dot{s}_{\alpha}
&= -\gamma \sum_{i=1}^n (s_\alpha-v_i)
= -\gamma\Big(n s_\alpha-\sum_{i=1}^n v_i\Big), \\
\dot{v}_i
&= -\gamma
\Big(
N v_i
- \sum_{\alpha=1}^{(1-r)N} u_{i\alpha}
- \sum_{\alpha=1}^{rN} s_{\alpha}
\Big),
\end{aligned}
\label{eq:frustrated-dynamics-noproj}
\end{equation}
where overdots denote derivatives with respect to time.

\noindent
\textbf{Fixed-point structure.}
The fixed-points of the gradient-flow dynamics are obtained by setting all
time derivatives in Eq.~\eqref{eq:frustrated-dynamics-noproj} to zero.
At any fixed point, all class-specific samples satisfy
$u_{i\alpha}=v_i$, while all shared samples take a common value
$s_\alpha=\bar{v}\equiv \frac{1}{n}\sum_i v_i$.
Substituting these relations back into the stationarity conditions yields
\begin{equation}
v_i = (1-r) v_i + r \bar{v} ,
\end{equation}
which implies
\begin{equation}
v_1 = v_2 = \cdots = v_n = \bar{v} .
\end{equation}
Thus, all label embeddings coincide at the fixed point, and the class representations collapse onto a single
shared value.
This collapse is a direct consequence of frustration and holds independently of the optimization path taken to reach the fixed point.
This behavior is not surprising, as frustration effectively introduces attractive interactions between label embeddings through the shared samples.
In the following, we analyze the dynamics by which the system approaches this collapsed fixed-point configuration.

\noindent
\textbf{Dynamics and time scales.}
The full gradient-flow dynamics Eq.~\eqref{eq:frustrated-dynamics-noproj} admits an exact decomposition into three invariant sectors.
Each sector is spanned by a set of symmetry-adapted effective degrees of freedom (DOF), evolves autonomously under a linear system, and is characterized by a distinct set of eigenvalues.

\paragraph{Sector I: sample-level fluctuations.}
For each class $i$, define the within-class mean and residuals
\begin{equation}
\bar u_i \equiv \frac{1}{(1-r)N}\sum_{\alpha=1}^{(1-r)N} u_{i\alpha},
\qquad
\delta u_{i\alpha} \equiv u_{i\alpha}-\bar u_i,
\end{equation}
with $\sum_\alpha \delta u_{i\alpha}=0$.
For the shared samples, define
\begin{equation}
\bar s \equiv \frac{1}{rN}\sum_{\alpha=1}^{rN}s_\alpha,
\qquad
\delta s_\alpha \equiv s_\alpha-\bar s,
\end{equation}
with $\sum_\alpha \delta s_\alpha=0$.
Sector~I is spanned by the residual DOF
$\{\delta u_{i\alpha}\}$ and $\{\delta s_\alpha\}$.
Subtracting the mean equations from the full dynamics yields
\begin{equation}
\dot{\delta u}_{i\alpha} = -\gamma \delta u_{i\alpha},
\qquad
\dot{\delta s}_\alpha = -\gamma n \delta s_\alpha.
\end{equation}
Thus, sector~I has eigenvalues
\begin{equation}
\lambda = -\gamma,
\qquad
\lambda = -\gamma n,
\end{equation}
corresponding to dispersion within class-specific and shared samples,
respectively.

\paragraph{Sector II: class-level deviations.}
Define global means across classes
\begin{equation}
\bar{u} \equiv \frac{1}{n}\sum_{i=1}^n \bar u_i,
\qquad
\bar{v} \equiv \frac{1}{n}\sum_{i=1}^n v_i,
\end{equation}
and class-deviation variables
\begin{equation}
\delta u_i \equiv \bar u_i-\bar{u},
\qquad
\delta v_i \equiv v_i-\bar{v},
\end{equation}
which satisfy $\sum_i \delta u_i=\sum_i \delta v_i=0$.
Sector~II is spanned by $\{\delta u_i,\delta v_i\}$ and captures inter-class
contrasts. For each independent deviation direction, the dynamics are
\begin{equation}
\frac{\text{d}}{\text{d}t}
\begin{pmatrix}
\delta u_i \\
\delta v_i
\end{pmatrix}
=
\gamma
\begin{pmatrix}
-1 & 1 \\
N(1-r) & -N
\end{pmatrix}
\begin{pmatrix}
\delta u_i \\
\delta v_i
\end{pmatrix}.
\end{equation}
The corresponding eigenvalues are
\begin{equation}
\lambda^{\mathrm{(II)}}_{\pm}
=
-\frac{\gamma}{2}
\Big[
N+1
\pm
\sqrt{(N+1)^2-4Nr}
\Big].
\end{equation}

\paragraph{Sector III: global mean sector.}
Sector~III is spanned by the fully symmetric DOF
\begin{equation}
(\bar{u},\;\bar{v},\;\bar s).
\end{equation}
Their dynamics are governed by
\begin{equation}
\frac{\text{d}}{\text{d}t}
\begin{pmatrix}
\bar{u} \\
\bar{v} \\
\bar s
\end{pmatrix}
=
\gamma
\begin{pmatrix}
-1 & 1 & 0 \\
N(1-r) & -N & Nr \\
0 & n & -n
\end{pmatrix}
\begin{pmatrix}
\bar{u} \\
\bar{v} \\
\bar s
\end{pmatrix}.
\end{equation}
This subsystem has one neutral eigenvalue
\begin{equation}
\lambda^{\mathrm{(III)}}_0 = 0,
\end{equation}
reflecting global translation symmetry, and two decaying eigenvalues
\begin{equation}
\lambda^{\mathrm{(III)}}_{\pm}
=
-\frac{\gamma}{2}
\Big[
N+n+1
\pm
\sqrt{(N-n+1)^2+4Nr(n-1)}
\Big].
\end{equation}
The decomposition shows that, at the level of the linearized dynamics,
the system admits several relaxation rates associated with different symmetry
sectors.
Not all of these modes, however, are relevant for classification or for the
observed training dynamics.

The global mean sector (Sector~III) governs only the relaxation of fully symmetric
degrees of freedom and leaves relative class geometry unchanged.
As a result, its modes do not manifest in the classification performance.
The relevant dynamics are therefore contained in Sectors~I and~II.
Sector~I captures sample-level fluctuations, with a rate of order $\gamma$
governing sample-to-class alignment and a rate of order $\gamma n$ associated
with internal relaxation among shared, frustrated samples.
The latter affects only relative positions within the frustrated subset and does
not directly contribute to class discrimination.
Sector~II captures class-level deviations and controls the evolution of
inter-class structure.
In the classification regime of interest, the number of samples per class $N$ is
large.
In this limit, the two eigenvalues of Sector~II separate into a
fast mode scaling as $\gamma N$ and a slower mode scaling as $\gamma r$.
The characteristic time scale associated with the fast mode, $1/(\gamma N)$,
vanishes as $N \to \infty$ and is therefore not observable in practice.

\begin{figure}[t!]
    \centering
    \includegraphics[width=0.8\linewidth]{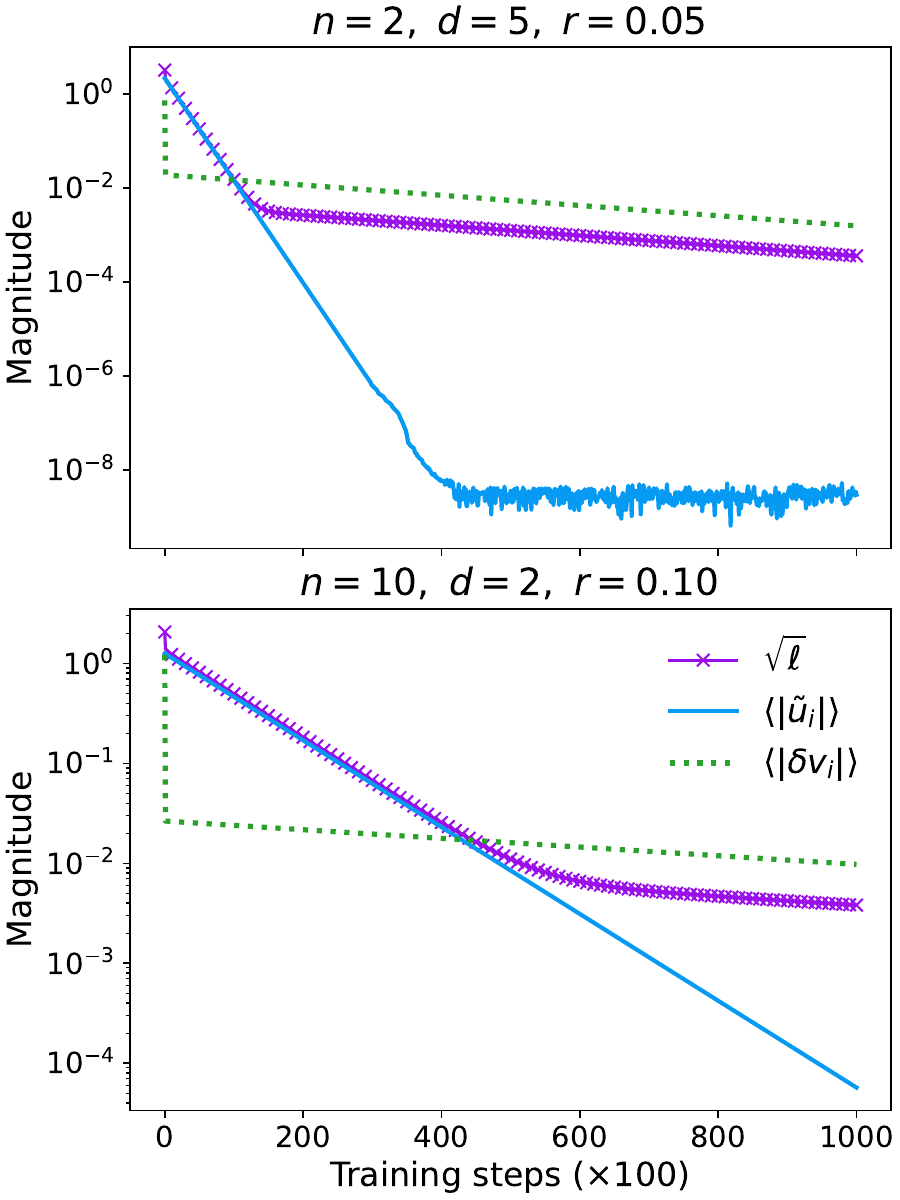}
   \caption{
    Training dynamics of the frustrated representation model, showing a clear
    separation of time scales.
    Curves show the square root of the loss $\sqrt{\ell}$, the mean magnitude
    of sample-level deviations $\langle |\delta u_i| \rangle$, and class-level
    deviations $\langle |\delta v_i| \rangle$ on a logarithmic scale.
    Top: $n=2$, $d=5$, $r=0.05$.
    Bottom: $n=10$, $d=2$, $r=0.10$.
    All experiments use $N=1000$ and a learning rate of $0.5$.
    Fast sample-level relaxation is followed by slow, frustration-controlled
    inter-class dynamics.
    }
    \label{fig:two_timescale_dynamics}
\end{figure}

Consequently, the effective learning dynamics reduce to two distinct time scales.
The first is the sample-to-representation alignment rate $\gamma$, which governs the rapid fitting of individual samples.
The second is the frustration-controlled rate $\gamma r$, which sets the slow, collective time scale associated with inter-class coupling and representation collapse.

\noindent
\textbf{Empirical results.}
Fig.~\ref{fig:two_timescale_dynamics} shows two representative examples of the
training dynamics in the frustrated representation model.
In both cases, the number of samples per class is fixed at $N=1000$, while the
number of classes $n$, the embedding dimension $d$, and the frustration level $r$
are varied, as indicated in the figure.
Despite these differences, both examples exhibit the same qualitative behavior.
We plot the square root of the loss, $\sqrt{\ell}$, since the loss is a
mean-squared error and taking the square root restores the natural distance scale
in representation space.
The loss dynamics separate into two stages, with an initial fast decay
corresponding to rapid sample-to-class alignment, followed by a much slower
relaxation driven by frustration-induced inter-class coupling, during which
class representations gradually approach one another.

\begin{figure}[t]
    \centering
    \includegraphics[width=\linewidth]{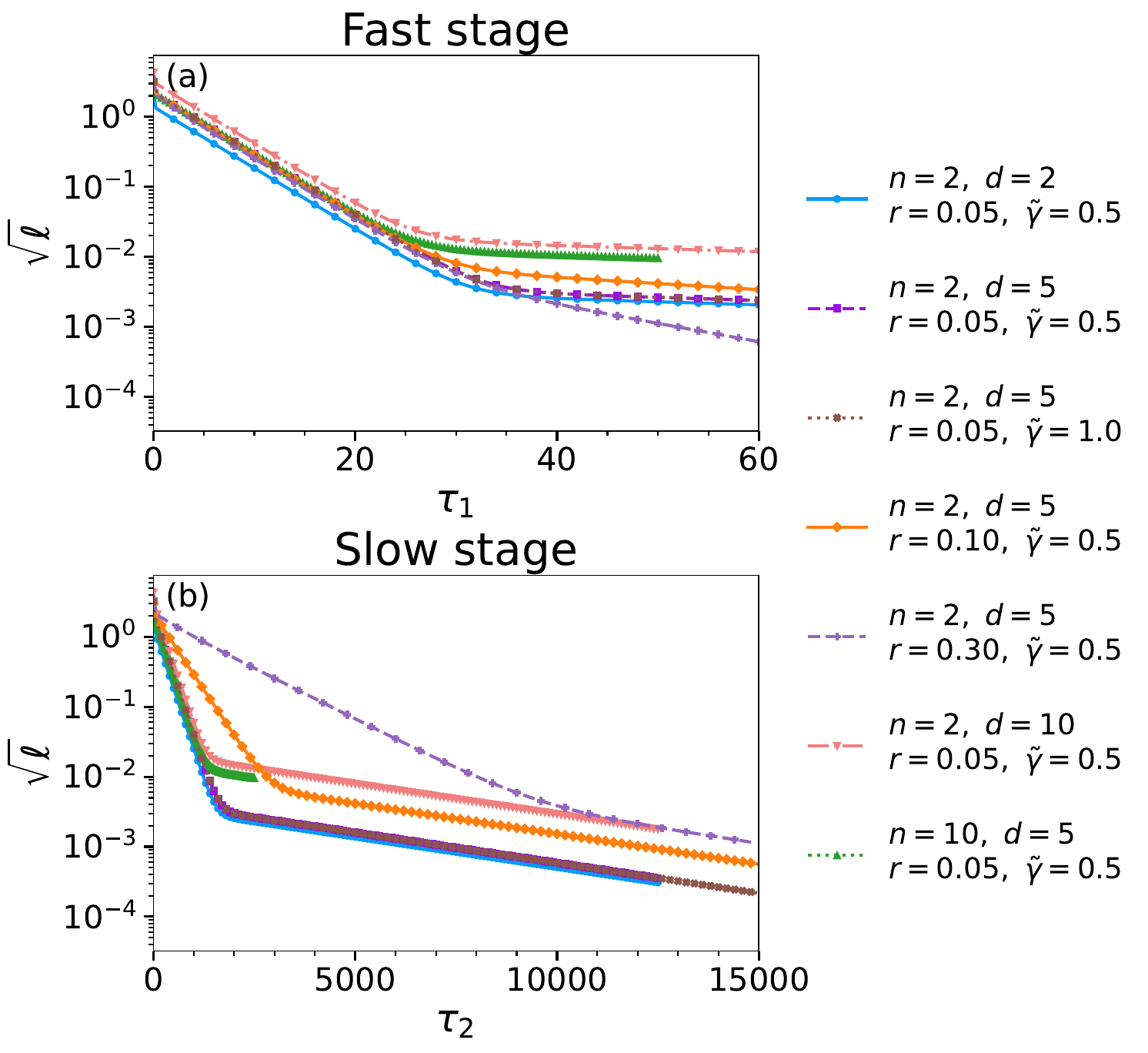}
   \caption{
    Two-stage relaxation of the training loss in the frustrated representation model.
    We plot the square root of the mean-squared loss, $\sqrt{\ell}$, using the fast rescaled time $\tau_1$ in (a) and the
    slow rescaled time $\tau_2$ in (b).
    The early-time decay in (a) reflects rapid sample-to-class alignment, while the long-time relaxation in (b) is
    controlled by frustration-induced inter-class coupling.
    Curves correspond to different choices of the number of classes $n$, embedding dimension $d$, frustration level $r$,
    and effective learning rate $\tilde{\gamma}$, as indicated in the legend.
    }
    \label{fig:two_timescale_loss_collapse}
\end{figure}

The two stages of the loss dynamics are naturally captured by two rescaled time
variables.
We define the fast time scale as $\tau_1 = \gamma t$, which governs the dynamics
of sample-to-class alignment, and the slow time scale as $\tau_2 = \gamma r t$,
which incorporates the effect of frustration through the fraction of shared
samples.
Fig.~\ref{fig:two_timescale_loss_collapse} shows the loss plotted against these rescaled times.
In panel~(a), the early-time decay collapses when expressed in terms of $\tau_1$,
indicating that this stage is dominated by a universal alignment process.
In panel~(b), the long-time decay collapses under $\tau_2$, revealing a distinct
frustration-controlled regime.
This second stage corresponds to the gradual collapse of class representations,
during which inter-class distinctions are reduced under competing alignment
constraints.

These results show that the observed two-stage behavior is a direct consequence
of a separation of relaxation time scales in the frustrated
representation dynamics.
The fast mode governs rapid sample-to-class alignment and corresponds to the
regime in which model performance improves.
At longer times, a much slower, frustration-controlled mode drives the gradual
collapse of class representations.
This separation provides a unified explanation for the empirical observation
that training accuracy can initially increase before eventually degrading as
collapse develops, as shown in Fig.~\ref{fig:mnist_cifar_cl}.


\section{
Projection and Stop-Gradient Prevent Representation Collapse Under Frustration
}
\label{sec:simsiam_minimal}
In this section, we study how the projection head and the stop-gradient mechanism help prevent collapse in the minimal frustrated model introduced above. We assume that the data and target branches share a common projection head. The difference between the cases with and without stop-gradient is shown in Fig.~\ref{fig:overview}(e) and \ref{fig:overview}(f). For simplicity, we consider a linear projection without a bias term, which can therefore be represented by a weight matrix $W$. The loss function can then be written as

\begin{equation} \label{eq:simsiam_minimal}
\begin{aligned}
\ell =
\frac{1}{2}\sum_{i=1}^n\sum_{\alpha=1}^{(1-r)N}
\left[(Wu_{i\alpha}-v_i)^2 + (Wv_i-u_{i\alpha})^2\right]\\
+
\frac{1}{2}\sum_{i=1}^n\sum_{\alpha=1}^{rN}
\left[(Ws_\alpha-v_i)^2 + (Wv_i-s_\alpha)^2\right]
\end{aligned}.
\end{equation}
Here, we do not explicitly annotate the stop-gradient operation in the loss function. Instead, its effect is made explicit at the level of the gradient-flow equations, which are given by

\begin{widetext}
\begin{equation}
\begin{aligned}
\dot{W}
&=
-\gamma \sum_{i=1}^n\sum_{\alpha=1}^{(1-r)N}
\left[(Wu_{i\alpha}-v_i)u_{i\alpha}^T + (Wv_i-u_{i\alpha})v_i^T\right]
-\gamma \sum_{i=1}^n\sum_{\alpha=1}^{rN}
\left[(Ws_\alpha-v_i)s_\alpha^T + (Wv_i-s_\alpha)v_i^T\right], \\[1.2ex]
\dot{u}_{i\alpha}
&=
-\gamma W^T(Wu_{i\alpha}-v_i)
\textcolor{accentblue}{-\gamma(u_{i\alpha}-Wv_i)}, \\
\dot{s}_\alpha
&=
-\gamma \sum_i W^T(Ws_\alpha-v_i)
\textcolor{accentblue}{-\gamma\sum_i(s_\alpha-Wv_i)}, \\
\dot{v}_i
&=
-\gamma W^T\!\left(
NWv_i
- \sum_{\alpha=1}^{rN}s_\alpha
- \sum_{\alpha=1}^{(1-r)N}u_{i\alpha}
\right)
\textcolor{accentblue}{-\gamma\!\left(
Nv_i
- \sum_{\alpha=1}^{rN}Ws_\alpha
- \sum_{\alpha=1}^{(1-r)N}Wu_{i\alpha}
\right)} .
\end{aligned}
\end{equation}
\end{widetext}
The terms highlighted in color blue are those that vanish when stop-gradient is applied. They arise solely from the absence of stop-gradient and correspond to additional reciprocal feedback between the data and label embeddings. Throughout this section, we further assume that $W$ has no zero eigenvalues. If $W$ possesses zero modes, the dynamics project out the corresponding eigenspaces, so that the effective degrees of freedom evolve only within the nonzero eigensubspace of $W$. This induces an effective dimensional reduction that we do not consider here.

\subsection{
Fixed-Point Structures Induced by Projected Heads, With vs. Without Stop-Gradient
}
\label{sec:simsiam fixed point}
We characterize the fixed-point manifold by imposing stationarity, i.e., setting all time derivatives to zero.
These stationary configurations govern the long-time behavior and therefore determine the system's final state.
We begin with the stop-gradient scenario, in which the stop-gradient operation eliminates certain terms and simplifies the dynamics substantially.

\noindent
\textbf{With stop-gradient.}
We start with the embedding sector of the fixed-point equations.
Since $W$ has no zero modes, they can be simplified as
\begin{equation}
\begin{aligned}
Wu_{i\alpha} &= v_i,\\
Ws_\alpha &= \bar v=\frac{1}{n}\sum_{j=1}^{n}v_j,\\
NWv_i &=  \sum_{\alpha=1}^{rN}s_\alpha
+ \sum_{\alpha=1}^{(1-r)N}u_{i\alpha}.
\end{aligned}
\label{eq:fp-stopgrad}
\end{equation}
Substituting Eq.~\eqref{eq:fp-stopgrad} into $\dot W = 0$ shows that the contributions involving $u_{i\alpha}$ and $s_\alpha$ vanish separately, with the latter dropping out because $\sum_i (\bar v - v_i)=0$. The remaining terms then cancel exactly by the global balance condition in the third line of Eq.~\eqref{eq:fp-stopgrad}. The stationarity condition for the projection matrix $W$ therefore imposes no additional constraint, and it is sufficient to analyze the embedding-sector equations alone. Acting with $W$ on the last equation in \eqref{eq:fp-stopgrad} and using the first two equations then yields a closed operator identity relating the label embeddings to $W^2$:
\begin{equation}
\big[W^2-(1-r)I\big]v_i
= r\bar v.
\label{eq:fp-stopgrad-compact}
\end{equation}
This identity restricts the spectrum of $W^2$ to at most two distinct eigenvalues within the subspace spanned by the label embeddings: $\lambda_{0}=1$ and $\lambda_{r}=1-r$. Let $\Pi_{0}$ and $\Pi_{r}$ denote the projection operators onto the corresponding eigenspaces. Projecting Eq.~\eqref{eq:fp-stopgrad-compact} onto these two sectors decouples the constraint into the following relations:
\begin{equation}
\Pi_{0} v_i = \Pi_{0}\bar v,\qquad \Pi_{r}\bar v = 0.
\end{equation}
The first relation enforces a geometric collapse within the $\lambda_{0}$ sector, requiring individual embeddings to coincide with their global mean. The second relation demonstrates that any non-collapsed components are strictly confined to the $\lambda_{r}$ sector and are admissible only under the condition that the global mean has no projection onto this subspace, $\Pi_{r}\bar v=0$. The existence of the $\lambda_r$ sector suggests that non-collapsed fixed points can emerge as viable physical solutions. In this regime, the system avoids a total loss of dimensionality. While the $\lambda_0$ component of the embeddings is forced into a singular configuration, the $\lambda_r$ component provides a manifold where class-wise separation can be sustained. Such solutions are permissible only when the embeddings are "centered" within this subspace, such that their collective centroid vanishes.

\noindent
\textbf{Without stop-gradient.}
The scenario without stop-gradient is more involved due to the additional terms. Without stop-gradient, the terms that would otherwise be removed remain coupled to the projection dynamics. As a result, the embedding-sector stationarity conditions are modified, and, as we show below, they exclude any non-collapsed fixed point for $r > 0$.
As before, we begin with the embedding-sector stationarity conditions, which can be simplified to
\begin{equation}
\begin{aligned}
    (W^TW+I)u_{i\alpha} &= (W^T+W)v_i\\
    (W^TW+I)s_\alpha & = (W^T+W)\bar{v}\\
    (W^TW+I)v_i &= \frac{1}{N}(W^T+W)\left(\sum_{\alpha=1}^{rN}s_\alpha
    + \sum_{\alpha=1}^{(1-r)N}u_{i\alpha}\right)
\end{aligned}.
\label{eq:fixed point no stop grad}
\end{equation}
Unlike the stop-gradient case, solutions of these embedding-sector conditions do not automatically satisfy the stationarity condition for the projection $W$, and thus impose additional constraints. 
For now, we focus on the embedding sector: substituting the first two equations into the last one gives
\begin{equation}
    (W^TW+I)v_i = B\left[r\bar{v} + (1-r)v_i\right],
\label{eq:no stop-gradient vi}
\end{equation}
where $B=(W^T+W)(W^TW+I)^{-1}(W^T+W)$ and $W^TW+I$ is positive definite and invertible.
Averaging Eq.~\eqref{eq:no stop-gradient vi} over $i$ and subtracting the mean relation yields the reduced condition
\begin{equation}
    (W^TW+I)^{1/2}\left[I-(1-r)C^2\right](W^TW+I)^{1/2}\delta v_i = 0,
\label{eq:vi condition no stop gradient}
\end{equation}
where $C = (W^TW+I)^{-1/2}(W^T+W)(W^TW+I)^{-1/2}$ and $\delta v_i = v_i - \bar{v}$. 
Therefore, to avoid collapse (i.e., to allow $\delta v_i\neq 0$), the operator $C^2$ must have an eigenvalue equal to $1/(1-r)$. Since $r > 0$, this required eigenvalue is strictly greater than $1$.
However, we can bound the spectrum of $C$ directly. Let $\lambda$ be an eigenvalue of $C$ with eigenvector $x$, such that $Cx = \lambda x$. By defining $v = (W^TW+I)^{-1/2}x$ and substituting the definition of $C$, we obtain the generalized eigenvalue equation
\begin{equation}
    (W^T+W)v = \lambda (W^TW+I)v.
\end{equation}
Left-multiplying this equation by $v^T$ yields $2v^TWv = \lambda (\|Wv\|^2 + \|v\|^2)$. We can then bound the magnitude of $\lambda$ as follows:
\begin{equation}
    |\lambda| = \frac{|2v^TWv|}{\|Wv\|^2 + \|v\|^2} \le \frac{2\|v\|\|Wv\|}{\|Wv\|^2 + \|v\|^2} \le 1.
\end{equation}
This demonstrates that all eigenvalues of $C$ fall within $[-1, 1]$, meaning $C^2$ cannot possibly admit an eigenvalue of $1/(1-r) > 1$. Consequently, the operator $I - (1-r)C^2$ is strictly positive definite. Eq.~\eqref{eq:vi condition no stop gradient} therefore uniquely enforces $\delta v_i = 0$. As a result, the geometric constraints of the embedding sector alone intrinsically prohibit any non-collapsed solutions, rendering a non-trivial difference sector impossible.

\noindent
\textbf{Discussion.}
The fixed-point structure shows that introducing the projection $W$ in the SimSiam spirit enlarges the dynamical system, but does not by itself resolve the collapse mechanism. In the fully coupled dynamics, the additional terms that remain without stop-gradient tighten the embedding-sector constraints. Concretely, Eq.~\eqref{eq:vi condition no stop gradient} shows that sustaining fluctuations $\delta v_i \neq 0$ would require a nontrivial spectral sector incompatible with the bound on the interaction operator, so the embedding-sector equations force $\delta v_i = 0$ for any $r>0$. The consequence is that the fully collapsed configuration remains the unique stationary state.

Stop-gradient, in contrast, alters the feedback structure between the two branches and removes the geometric obstruction that otherwise enforces collapse. The stationary constraints then separate into two spectral sectors of $W^2$: the $\lambda_0=1$ sector enforces collapse onto the mean, while the $\lambda_r=1-r$ sector can sustain nontrivial class-wise structure provided the embeddings are centered, $\Pi_r\bar v=0$. Thus, stop-gradient is the ingredient that stabilizes a non-collapsed fixed-point manifold, whereas the projection alone does not prevent collapse once the full coupling is retained.

\subsection{
Dynamics Under Projection and Stop-Gradient
}
\label{sec:simsiam-dynamics}

We briefly discuss the dynamics of the system. Since the introduction of the projection $W$,
the system is nonlinear and thus a closed-form analytical solution is not explicit.
Here, we derive self-consistent equations in the spirit of DMFT \cite{georges1996dynamical,sompolinsky1988chaos}.
As in Sec.~\ref{sec:frustrated dynamics}, we can decompose the dynamics into a mean sector, a class-deviation sector, and a within-class fluctuation sector. The difference is that each sector evolves in a time-dependent medium set by $W(t)$.
This yields a Dyson-type integral formulation with a memory kernel.

\noindent
\textbf{$W$-sector.}
We start with the projection sector. The dynamics of $W$ can be written as the linear matrix ODE
\begin{equation}
\dot W = -\gamma\Big(W S(t)-R(t)\Big),
\label{eq:W-linear}
\end{equation}
where
\begin{equation}
\begin{aligned}
S(t)
&= \sum_{i=1}^n\Big[(1-r)N \bar u_i\bar u_i^T +   C_{i,u}(t) + N v_i v_i^T\Big]
\\
&\quad\qquad\qquad\qquad\qquad + n\Big(rN \bar s\bar s^T + C_s(t)\Big),
\\[0.5ex]
R(t)
&= \sum_{i=1}^n\Big[(1-r)N (v_i\bar u_i^T+\bar u_i v_i^T)
\\
&\quad\qquad\qquad\qquad\qquad\quad
+ rN (v_i\bar s^T+\bar s v_i^T)\Big].
\end{aligned}
\label{eq:R-general}
\end{equation}
Here $  C_{i,u}(t)$ and $C_s(t)$ are the within-group covariances, defined as
\begin{equation}
\begin{aligned}
 C_{i,u}(t)&=\sum_{\alpha=1}^{(1-r)N}\big[u_{i\alpha}(t)-\bar u_i(t)\big]\big[u_{i\alpha}(t)-\bar u_i(t)\big]^T,
\\
C_s(t)&=\sum_{\alpha=1}^{rN}\big[s_{\alpha}(t)-\bar s(t)\big]\big[s_{\alpha}(t)-\bar s(t)\big]^T.
\end{aligned}
\label{eq:cov-def}
\end{equation}
The solution of \eqref{eq:W-linear} admits the integral representation
\begin{align}
W(t)
&= W(0) \Phi(t,0) + \gamma\int_0^t R(\tau)\Phi(t,\tau)d\tau,
\label{eq:W-VoC}
\end{align}
where the propagator $\Phi(t,\tau)$ is defined by
\begin{align}
\partial_t\Phi(t,\tau) &= -\gamma\Phi(t,\tau)S(t),
\qquad
\Phi(\tau,\tau)=I.
\label{eq:Phi}
\end{align}
Here $R(t)$ collects the mixed second moments between the projection outputs and the label embeddings.
It is not an external input. Instead, $R(t)$, and similarly $S(t)$, is determined self-consistently by the embedding
trajectories through \eqref{eq:R-general} together with the embedding-sector dynamics derived below.

\noindent
\textbf{Embedding sector.}
With the projection $W(t)$, the embeddings evolve in a time-dependent medium set by the Gram operator
\begin{equation}
M(t)\equiv W(t)^T W(t).
\label{eq:M-def}
\end{equation}
This operator acts as an effective medium for the embedding dynamics, replacing the constant relaxation
rates of the unprojected model with matrix-valued couplings.
We follow the same sector decomposition as in Sec.~\ref{sec:frustrated dynamics}, separating the dynamics into the mean sector,
the class-deviation sector, and sample-level fluctuations, all evolving in the medium set by $M(t)$, with $M(t)$
ultimately determined by the coupled projection dynamics.

Under this decomposition, the mean sector evolves as
\begin{equation}
\begin{aligned}
\dot{\bar u}
&= -\gamma M \bar u + \gamma W^T\bar v,\\
\dot{\bar v}
&= -\gamma N M \bar v
+ \gamma N W^T \bigg[(1-r)\bar u + r\bar s\bigg],\\
\dot{\bar s}
&= -\gamma n M \bar s + \gamma n W^T\bar v,
\end{aligned}
\label{eq:mean-sector}
\end{equation}
the class-deviation modes satisfy
\begin{equation}
\begin{aligned}
\dot{\delta u}_i
&= -\gamma M \delta u_i + \gamma W^T\delta v_i,\\
\dot{\delta v}_i
&= -\gamma N M \delta v_i
+ \gamma N(1-r) W^T\delta u_i,
\end{aligned}
\label{eq:deviation-sector}
\end{equation}
and, away from the permutation-symmetric manifold, the sample-level fluctuations are summarized by the within-group
covariances $(  C_{i,u},C_s)$ obeying the anticommutator flow
\begin{equation}
\begin{aligned}
\dot   C_{i,u}(t) &= -\gamma \{M(t),  C_{i,u}(t)\},\\
\dot C_s(t) &= -\gamma n \{M(t),C_s(t)\},
\end{aligned}
\label{eq:covariance-sector}
\end{equation}
with $\{A,B\}=AB+BA$.

\begin{figure}[t!]
    \centering
    \includegraphics[width=0.98\linewidth]{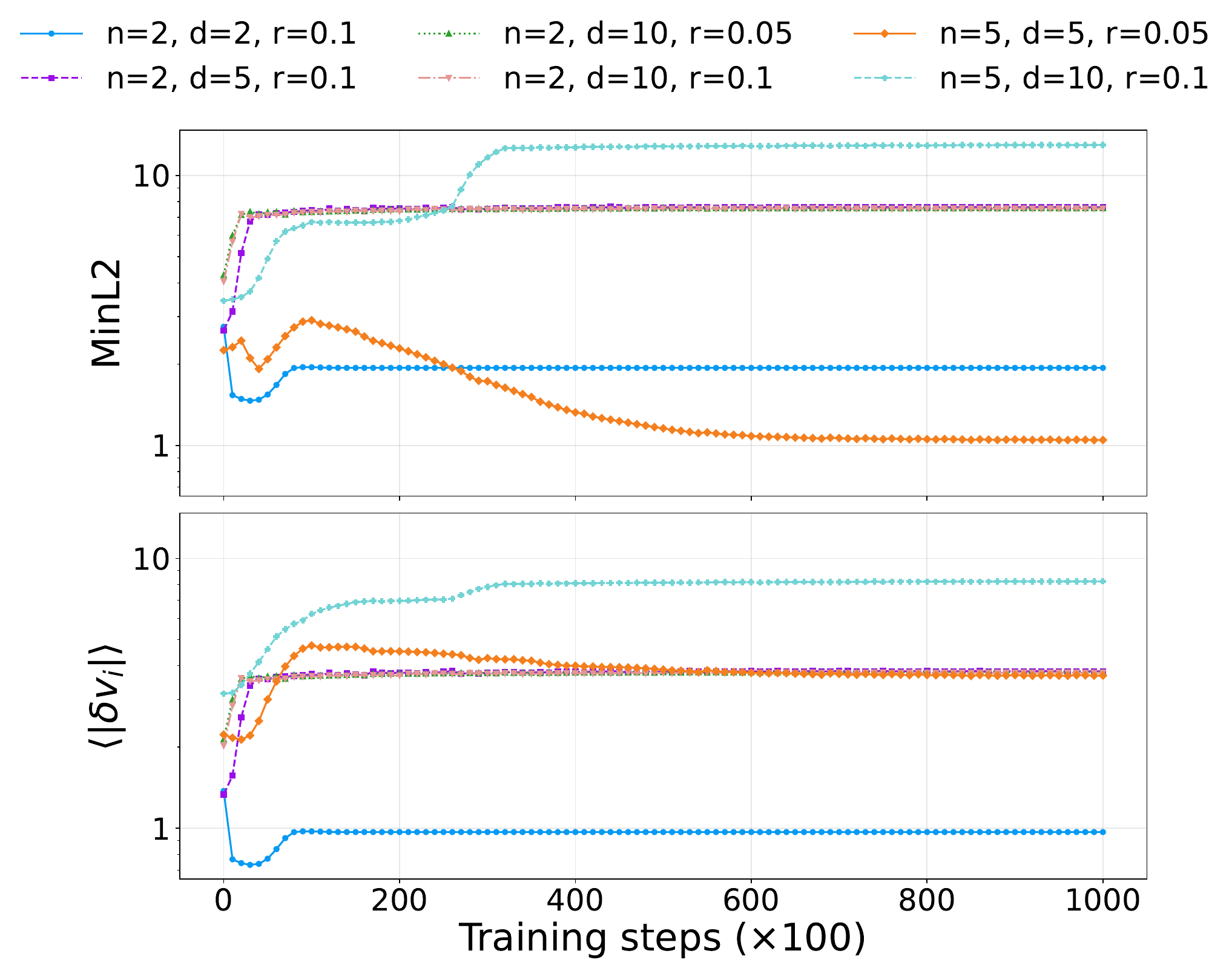}
    \caption{
    Training dynamics of the minimal model with projection head and stop-gradient, shown for different numbers of classes $n$, embedding dimensions $d$, and frustration ratios $r$ (legend).
    Top: the minimum pairwise distance between label embeddings, $\mathrm{MinL2}$.
    Bottom: the mean magnitude of label-embedding deviations, $\langle|\delta v_i|\rangle$.
    Both panels use a logarithmic $y$-axis and share the same $y$-range for direct comparison; the horizontal axis denotes training steps (in units of $10^2$).
    }
    \label{fig:with_stop_grad_dyn}
\end{figure}

We introduce the propagators generated by $M(t)$,
\begin{equation}
\begin{aligned}
\partial_t G_1(t,\tau) &= -\gamma M(t) G_1(t,\tau),
\qquad
G_1(\tau,\tau)=I,\\
\partial_t G_N(t,\tau) &= -\gamma N M(t) G_N(t,\tau),
\qquad
G_N(\tau,\tau)=I,\\
\partial_t G_n(t,\tau) &= -\gamma n M(t) G_n(t,\tau),
\qquad
G_n(\tau,\tau)=I.
\end{aligned}
\label{eq:greens}
\end{equation}
These propagators describe transport in a time-dependent effective medium $M(t)$, replacing the constant-rate exponential decay of the unprojected model. In terms of these propagators, the mean sector admits the integral representation

\begin{equation}
\begin{aligned}
\bar u(t)
&= G_1(t,0) \bar u(0)
+ \gamma\int_0^t G_1(t,\tau) W(\tau)^T\bar v(\tau) d\tau,
\\
\bar s(t)
&= G_n(t,0) \bar s(0)
+ \gamma n\int_0^t G_n(t,\tau) W(\tau)^T\bar v(\tau) d\tau,
\\
\bar v(t)
&= G_N(t,0) \bar v(0)
+ \gamma N\int_0^t G_N(t,\tau) W(\tau)^T
\\
&\qquad\qquad\qquad\qquad\times
\Big[(1-r)\bar u(\tau)+r\bar s(\tau)\Big] d\tau,
\end{aligned}
\label{eq:mean-integral}
\end{equation}
and the class-deviation sector is equivalently
\begin{equation}
\begin{aligned}
\delta u_i(t)
&= G_1(t,0) \delta u_i(0)
+ \gamma\int_0^t G_1(t,\tau) W(\tau)^T\delta v_i(\tau) d\tau,
\\
\delta v_i(t)
&= G_N(t,0) \delta v_i(0)
+ \gamma N(1-r)\int_0^t G_N(t,\tau)
\\
&\qquad\qquad\qquad\qquad\qquad\qquad\times W(\tau)^T
\delta u_i(\tau) d\tau.
\end{aligned}
\label{eq:dev-integral}
\end{equation}
At the sample level, the microscopic modes propagate as
\begin{equation}
\begin{aligned}
\delta u_{i\alpha}(t) &= G_1(t,0) \delta u_{i\alpha}(0),
\\
\delta s_{\alpha}(t) &= G_n(t,0) \delta s_{\alpha}(0),
\end{aligned}
\label{eq:micro-prop}
\end{equation}
which implies the covariance propagation laws
\begin{equation}
\begin{aligned}
  C_{i,u}(t) &= G_1(t,0)   C_{i,u}(0) G_1(t,0)^T,
\\
C_s(t) &= G_n(t,0) C_s(0) G_n(t,0)^T.
\end{aligned}
\label{eq:cov-prop}
\end{equation}
The integral representations \eqref{eq:mean-integral}-\eqref{eq:cov-prop} make explicit how the embedding degrees of freedom
are transported in the effective medium $M(t)$, while the projection sector is encoded by the propagator $\Phi(t,\tau)$ defined in
\eqref{eq:Phi} and the variation-of-constants formula \eqref{eq:W-VoC}.
Although all embedding modes evolve in the same medium, the three channels retain distinct characteristic rates set by
$\gamma$, $\gamma N$, and $\gamma n$ through $G_1$, $G_N$, and $G_n$, respectively.
The frustration ratio $r$ does not enter the propagators directly.
It appears in the source terms, for instance through the combination $(1-r)\bar u + r\bar s$ in \eqref{eq:mean-integral}.
It also enters indirectly through the learned projection, which determines $M(t)=W(t)^T W(t)$.
As a result, the frustration-induced attraction acts only through the projection sector.
Its effect on the embedding dynamics is mediated by $W(t)$ and can be reduced when $M(t)$ relaxes toward fixed points that suppress this coupling, as discussed in Sec.~\ref{sec:simsiam fixed point}.
Together, these relations provide a convenient starting point for further analytical reductions.
For example, if the projection and embedding sectors separate in time scales, one may derive effective dynamics by integrating out the fast sector.
In this way, the projection can be approximated as quasi-static while the embeddings follow a reduced evolution.
Conversely, one can obtain an effective equation for the slow drift of $W$ near the collapsed configuration.
We leave a detailed development of these reduced descriptions for future work and do not pursue it further here.

\subsection{Empirical Results}
\begin{figure}[t!]
    \centering
    \includegraphics[width=0.98\linewidth]{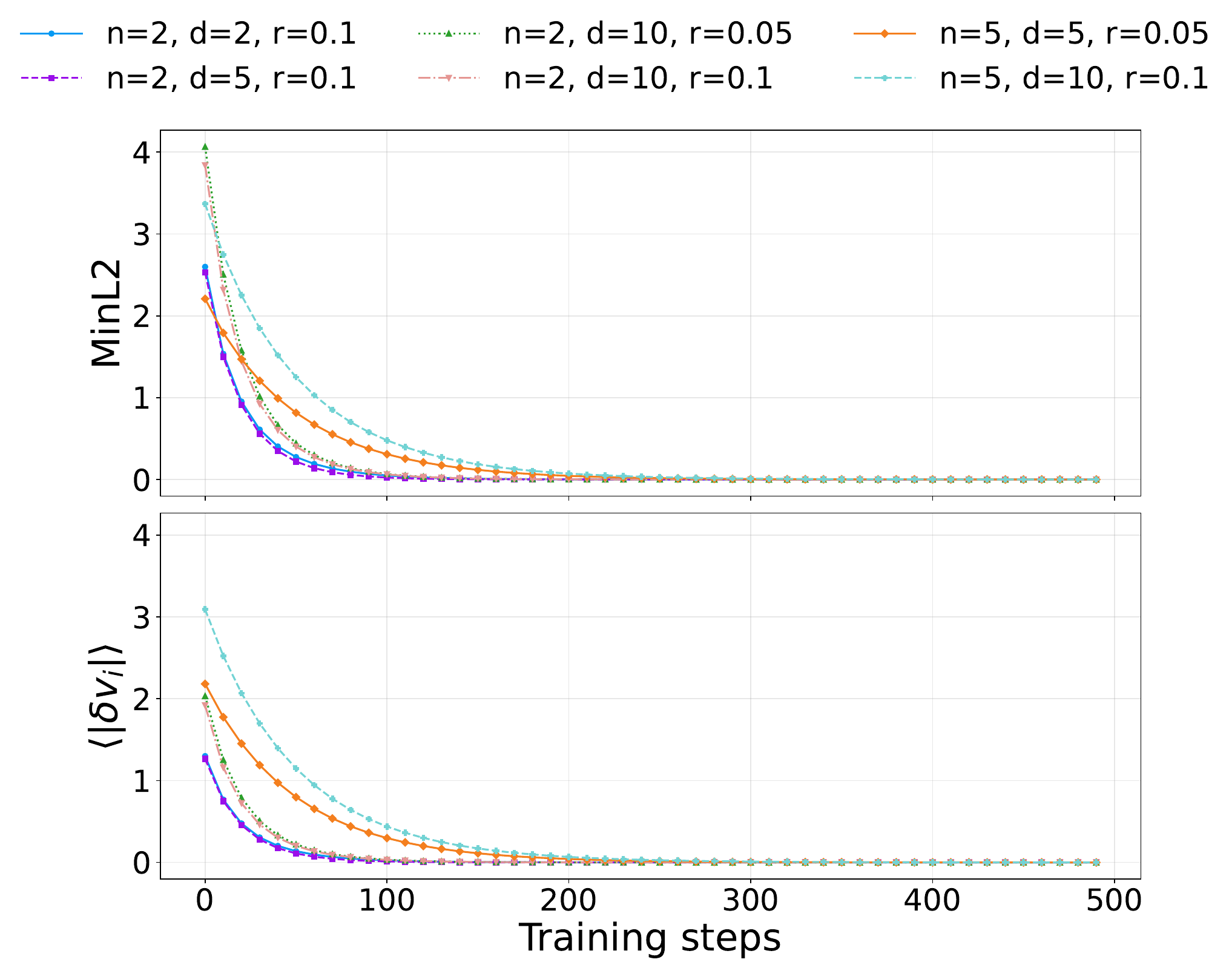}
    \caption{
    Training dynamics of the minimal model with projection head but \emph{without} stop-gradient, shown for different numbers of classes $n$, embedding dimensions $d$, and frustration ratios $r$ (legend).
    Top: the minimum pairwise distance between label embeddings, $\mathrm{MinL2}$.
    Bottom: the mean magnitude of label-embedding deviations, $\langle|\delta v_i|\rangle$.
    Both quantities rapidly decay toward zero, indicating convergence to a fully collapsed configuration in which label embeddings become indistinguishable.
    The horizontal axis denotes training steps.
    }
    \label{fig:without_stop_grad_dyn}
\end{figure}
We next validate our theoretical results with empirical training runs of the minimal model with projection head.
Figures~\ref{fig:with_stop_grad_dyn} and \ref{fig:without_stop_grad_dyn} compare the training dynamics with and without stop-gradient. In both figures we track two quantities. The first is the mean deviation magnitude $\langle|\delta v_i|\rangle$, which is the main object in our theoretical analysis and measures the overall spread of the label embeddings. The second is the minimum pairwise distance between label embeddings, $\mathrm{MinL2}=\min_{i\neq j}\|v_i-v_j\|_2$, which directly tests whether any two classes have become indistinguishable. While the theory is formulated in a class-symmetric setting, a single finite run need not realize the permutation symmetry between classes. The dynamics can therefore exhibit partial collapse in which only a subset of labels collapse onto one another. In that case $\langle|\delta v_i|\rangle$ can remain finite even though at least one pair has already collapsed, and $\mathrm{MinL2}$ provides a sensitive indicator of the onset of collapse.

With stop-gradient in Figure~\ref{fig:with_stop_grad_dyn}, $\langle|\delta v_i|\rangle$ decays rapidly after a short transient, while $\mathrm{MinL2}$ typically settles to a nonzero plateau whose value depends on the initialization as well as $(n,d,r)$. The label embeddings therefore remain separated rather than collapsing to a single point. Without stop-gradient in Figure~\ref{fig:without_stop_grad_dyn}, both $\mathrm{MinL2}$ and $\langle|\delta v_i|\rangle$ rapidly decay toward zero across all tested settings. The dynamics then converge to a fully collapsed configuration in which the label embeddings become indistinguishable, consistent with an attractive collapsed fixed point under full-gradient updates.

\begin{figure}[tb]
    \centering
    \includegraphics[width=0.98\linewidth]{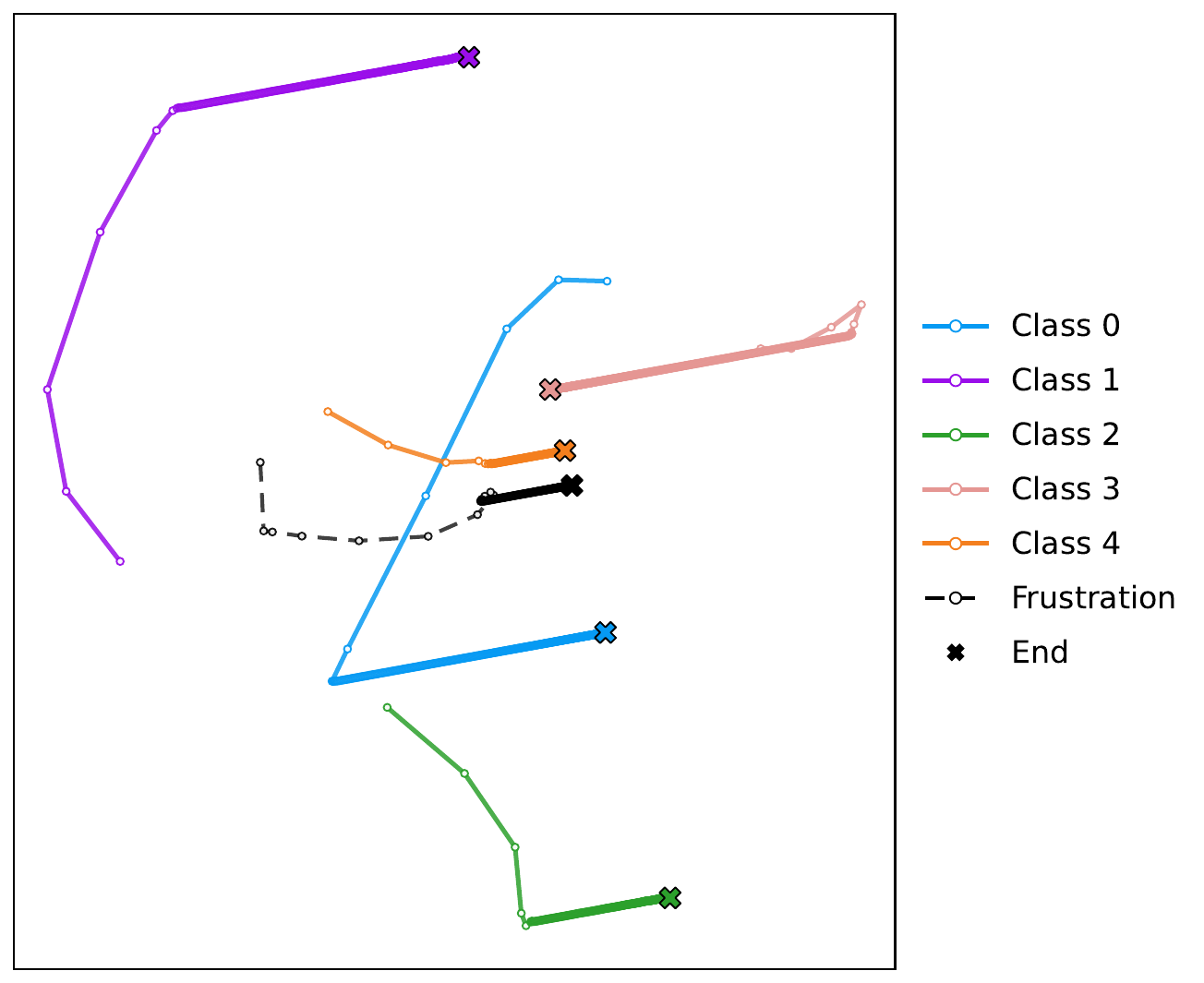}
    \caption{Schematic training trajectories in the minimal model with a single frustrated sample.
    Shown are the 2D paths of the projected class label embeddings $Wv_i$ for $n=5$ classes (colored curves) together with the frustration embedding $s$ (black dashed curve), with embedding dimension $d=2$ and $N=50$ samples per class. The single frustrated sample corresponds to a frustration ratio $r=1/50$. Positions are recorded every $1000$ optimization steps (open circles), and the final state is marked by a cross.}
    \label{fig:simsiam_non_collapse_example}
\end{figure}

To complement these aggregate curves, Figure~\ref{fig:simsiam_non_collapse_example} plots an illustrative two-dimensional trajectory of the projected label embeddings $Wv_i$, with one curve per class, together with the frustration embedding $s$ as training evolves. We observe a strongly anisotropic evolution in which the trajectories contract and align along one direction while remaining spread along a transverse direction, so that separation is lost only partially rather than uniformly. This pattern supports our theoretical picture in which collapse occurs within the eigenspace of $W^2$ with eigenvalue $1$, whereas non-collapse can persist in the eigenspace with eigenvalue $1-r$.

\begin{figure}[t]
    \centering
    \includegraphics[width=0.96\linewidth]{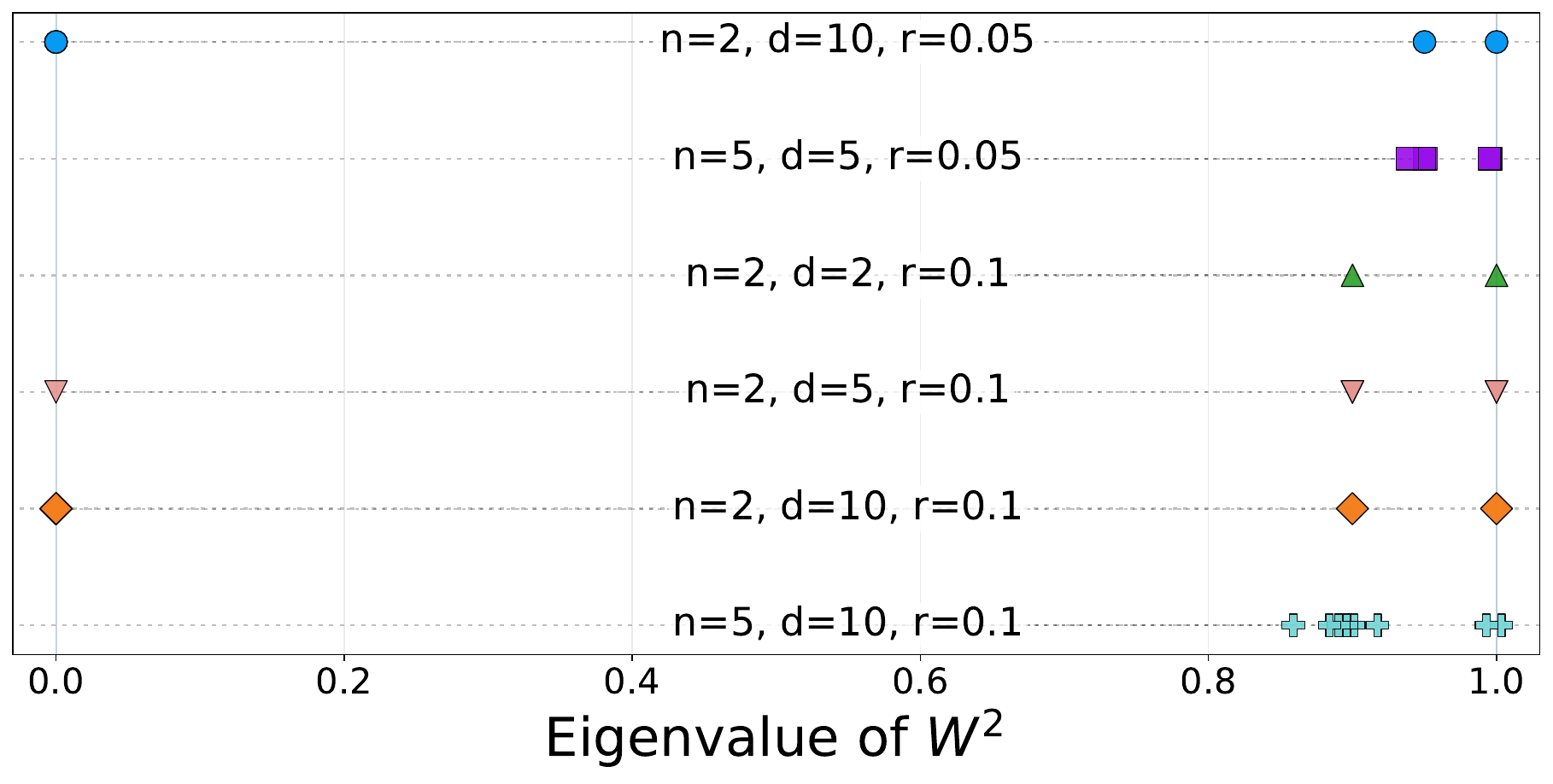}
    \caption{Eigenvalue spectra of $W^2$ at the end of training (100{,}000 steps) for the stop-gradient (SG) runs. Each row corresponds to one  setting $(n,d,r)$, and markers denote the eigenvalues of $W^2$. Across settings, the spectrum concentrates near a small number of well-separated values, with an eigenvalue near $0$ and the remaining eigenvalues grouped near $1$ and $1-r$.}
    \label{fig:eigenvalues_W2}
\end{figure}

Figure~\ref{fig:eigenvalues_W2} provides a spectral summary by plotting the eigenvalues of $W^2$ at the end of training for each stop-gradient run. The spectrum concentrates near a small number of well-separated values. For convenience, our theoretical analysis assumes that $W$ has no zero modes. In practice, some eigenvalues can appear near $0$, indicating that the dynamics effectively evolve in a lower-dimensional subspace, with collapse occurring along the corresponding directions. Apart from these modes, the spectra align with the theoretical prediction that the remaining eigenvalues cluster near two values, $1$ and $1-r$. The eigenvalue-$1$ sector identifies directions that undergo collapse, whereas the eigenvalue-$(1-r)$ sector identifies directions that remain non-collapsed.

\section{
Representation Collapse From Embedding-Wise Frustration to A Linear Teacher--Student Model
}
\label{sec:teacher-student}
We now move beyond the embedding-only minimal model to a parametrized setting, where representations are generated by a learned map from inputs rather than treated as free variables. We adopt a linear teacher-student construction because it provides a simple controlled setting: in a teacher-student model, a fixed teacher generates the target labels while a student model is trained to learn them. Here, the task is separable in the absence of corruption, while tunable label corruption gives direct control over the frustration ratio. This allows us to test whether the collapse mechanism identified in the embedding-only model persists once representations can be learned through a function class under controlled levels of frustration.

\subsection{Teacher Model and Data Generation}
We use a simple teacher--student construction in which the teacher is a linear binary classifier in $d$ dimensions.
The teacher assigns labels according to the sign of a fixed direction,
\begin{equation}
y(x)=\mathrm{sign}(w^{\top}x),
\end{equation}
where we take $d=64$ and choose $w$ to be the normalized alternating vector $(1,-1,1,-1,\dots)$.

\begin{figure}[t]
    \centering
    \includegraphics[width=\linewidth]{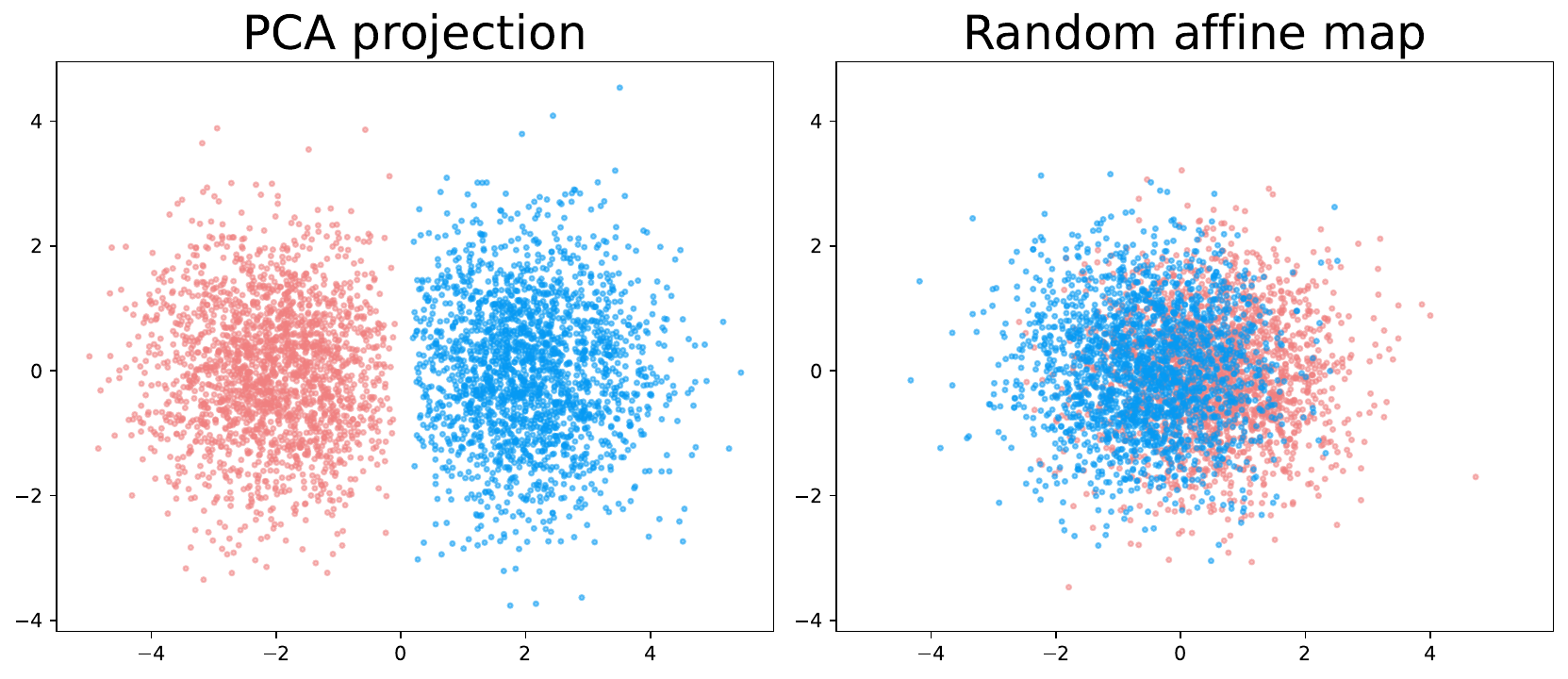}
    \caption{
    Two-dimensional visualizations of the synthetic teacher-student dataset in $d=64$, where different colors denote the two classes.
    Left: PCA projection. Right: random affine map to $\mathbb{R}^2$.
    }
    \label{fig:data plot}
\end{figure}

Given this teacher, we generate two Gaussian classes in $\mathbb{R}^{64}$ with isotropic noise and opposite mean shifts along $w$.
Concretely, we sample $x\sim \mathcal{N}(+2\mu, I)$ for class $+1$ and $x\sim \mathcal{N}(-2\mu, I)$ for class $-1$.
To avoid samples near the decision boundary, we discard points whose projection onto $\mu$ has magnitude smaller than $0.2$.
Figure~\ref{fig:data plot} shows two subfigures that visualize the same $64$-dimensional dataset.
The left subfigure shows a PCA projection that clearly separates the two classes and reflects the underlying linear separability.
The right subfigure shows a random affine map to $\mathbb{R}^2$ that can make the classes appear overlapping and indicates that a randomly initialized linear layer does not separate them.

To introduce frustration, given $N$ samples in each class, we choose $rN$ points from each class uniformly at random and overwrite their feature vectors by assigning them to the opposite class's peak, i.e., to $\pm 2\mu$, while keeping their original labels unchanged.
This yields a fraction $r$ of frustrated samples per class and keeps the dataset perfectly balanced.
Because the same operation is applied to both classes, it preserves the permutation symmetry between classes and isolates the effect of frustration from trivial class-imbalance effects.

\subsection{
Representation Collapse and Dynamics Under Frustration
}
We first analyze the collapse dynamics in a minimal linear setting.
We model the sample embeddings with a linear student $x \mapsto Wx+b$ and represent each class by a learned label embedding $v_i$, as shown in Fig.~\ref{fig:overview}(g).
Using the mean-squared error (MSE) objective, the loss is
\begin{equation}
    \ell = \sum_{i,\alpha}\big(Wx_{i\alpha} + b - v_i\big)^2,
\end{equation}
where $\alpha$ indexes samples within a class, $i\in\{1,2\}$ indexes the two classes, $v_i$ is the label embedding for class $i$, and $W$ and $b$ are the weights and bias of the student model. Before turning to empirical results, we briefly discuss two key differences between this setup and the pure embedding models.

\begin{figure}[t]
    \centering
    \includegraphics[width=\linewidth]{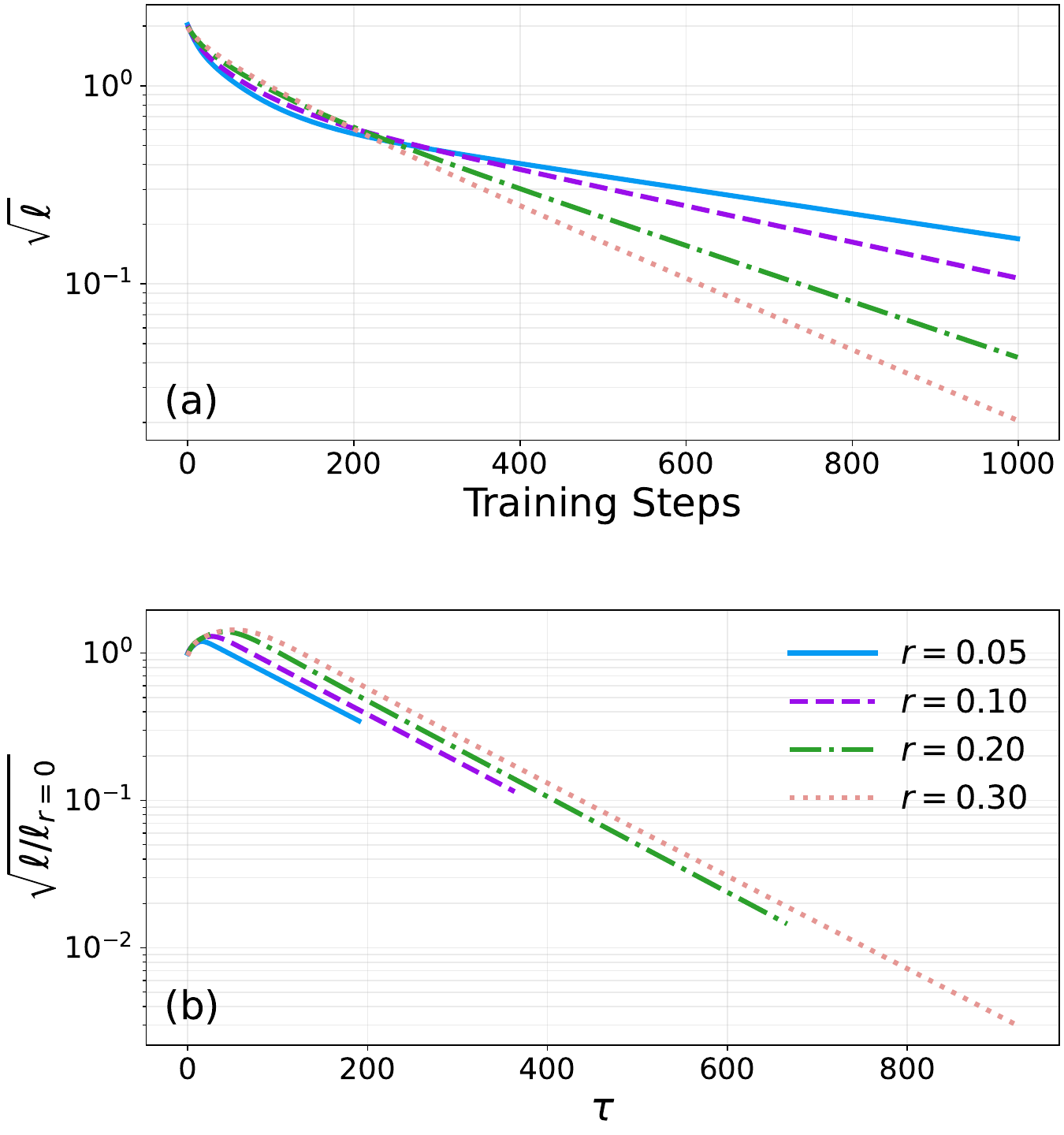}
    \caption{
    Training dynamics of the linear student-label embedding model with embedding dimension $5$ at learning rate $0.005$ for different frustration rates $r$.
    (a) $\sqrt{\ell}$ as a function of training steps on a log scale.
    (b) Rescaled loss ratio $\sqrt{\ell/\ell_{r=0}}$ versus the effective time $\tau$.
    }
    \label{fig:teacher student cl loss}
\end{figure}

\noindent
\textbf{Scaling of the loss $\ell$.}
This MSE objective has a simple scale symmetry.
If we multiply the entire representation by a constant factor $k$, meaning we rescale the student output and the label embeddings together,
\begin{equation}
    W \rightarrow kW,\qquad b \rightarrow kb,\qquad v_i \rightarrow kv_i,
\end{equation}
then every residual $Wx_{i\alpha}+b-v_i$ is multiplied by $k$.
Since the loss is a sum of squared residuals, it rescales as $\ell \rightarrow k^2\ell$.
This scaling direction becomes relevant when the student model cannot fit the targets perfectly, so the residuals $Wx_{i\alpha}+b-v_i$ cannot be driven to zero.
That is the generic situation here, because a single linear layer cannot, in general, align all sample embeddings $Wx_{i\alpha}+b$ with the corresponding label embeddings $v_i$.
As a result, gradient descent can keep reducing the loss by shrinking the overall scale of $(W,b,v_i)$, which preserves the relative embedding geometry and therefore leaves the decision structure essentially unchanged.
At the same time, this contraction decreases the absolute distances between label embeddings, driving them closer together.
In contrast, in the pure embedding case the model can drive $\ell$ to zero, so this scaling freedom plays no role.
Notably, the same rescaling mechanism is present even when $r=0$, so it does not rely on frustration.

\noindent
\textbf{Effective frustration rate.}
A second difference is the way frustration is implemented.
Rather than introducing universally frustrated samples that couple to all classes, we select a balanced set of samples from each class and overwrite their feature vectors by placing them at the opposite class peak, while keeping their labels unchanged.
In this construction, each frustrated sample acts as a mismatched constraint of strength comparable to two label-flips in the pure embedding setting, effectively amplifying the frustration.
It is therefore convenient to define an effective frustration rate
\begin{equation}
    r_{\mathrm{eff}} = \frac{2r}{1+r},
\end{equation}
and to express the relevant time scale in terms of $r_{\mathrm{eff}}$ rather than $r$.

\noindent
\textbf{Empirical results.}
We now turn to the empirical training dynamics in Fig.~\ref{fig:teacher student cl loss}. In these experiments, the linear student maps $64$-dimensional inputs to a $5$-dimensional embedding space.
Panel~(a) plots $\sqrt{\ell}$ versus training steps for several frustration rates $r$ and reveals a clear two-stage dynamics.
The loss first drops rapidly in an early transient.
This early-stage decay is not well described by a single exponential, reflecting the coupled evolution induced by the model structure even in this linear setting.
At late times, the dynamics cross over to a regime with an approximately exponential decay.
The separation of time scales is pronounced, and the duration of the initial transient decreases as $r$ increases.
To factor out the trivial decay shared by all runs, we normalize the loss by the no-frustration baseline and plot $\sqrt{\ell/\ell_{r=0}}$.
Panel~(b) shows this ratio as a function of the effective time $\tau=r_{\mathrm{eff}}t$.
After normalization, the curves nearly collapse in the early stage, indicating that the initial transient is largely set by the $r=0$ time scale.
At late times, the curves approach straight lines on the log scale, indicating an approximately exponential decay.
Moreover, these lines are nearly parallel across different $r$, which implies that the decay rates are the same once time is measured in $\tau=r_{\mathrm{eff}}t$.
This is consistent with our theory that frustration drives the late-time collapse and sets the relevant time scale through the (effective) frustration ratio.

\subsection{
Dynamics Under Projection and Stop-Gradient
}

Lastly, we study training in a linear teacher--student setting.
As shown in Fig.~\ref{fig:overview}(h), we focus on the variant with projection head and stop-gradient, which prevents complete collapse and supports stable non-collapsed fixed points.
For simplicity, we use a linear projection head without bias.
The loss function is then written as
\begin{equation} \label{eq:teacher_student_mse}
\begin{aligned}
    \ell = &\sum_{i\alpha}\Bigg(W'(Wx_{i\alpha} + b) -\text{sg}[v_i]\Bigg)^2 \\
    &+  \sum_{i\alpha}\Bigg(W'v_i - \text{sg}[Wx_{i\alpha} + b]\Bigg)^2,
\end{aligned}
\end{equation}
where $W'$ is the projection-head weight and $\text{sg}[\cdot]$ denotes the stop-gradient operator during backpropagation.

\begin{figure}[t]
    \centering
    \includegraphics[width=\linewidth]{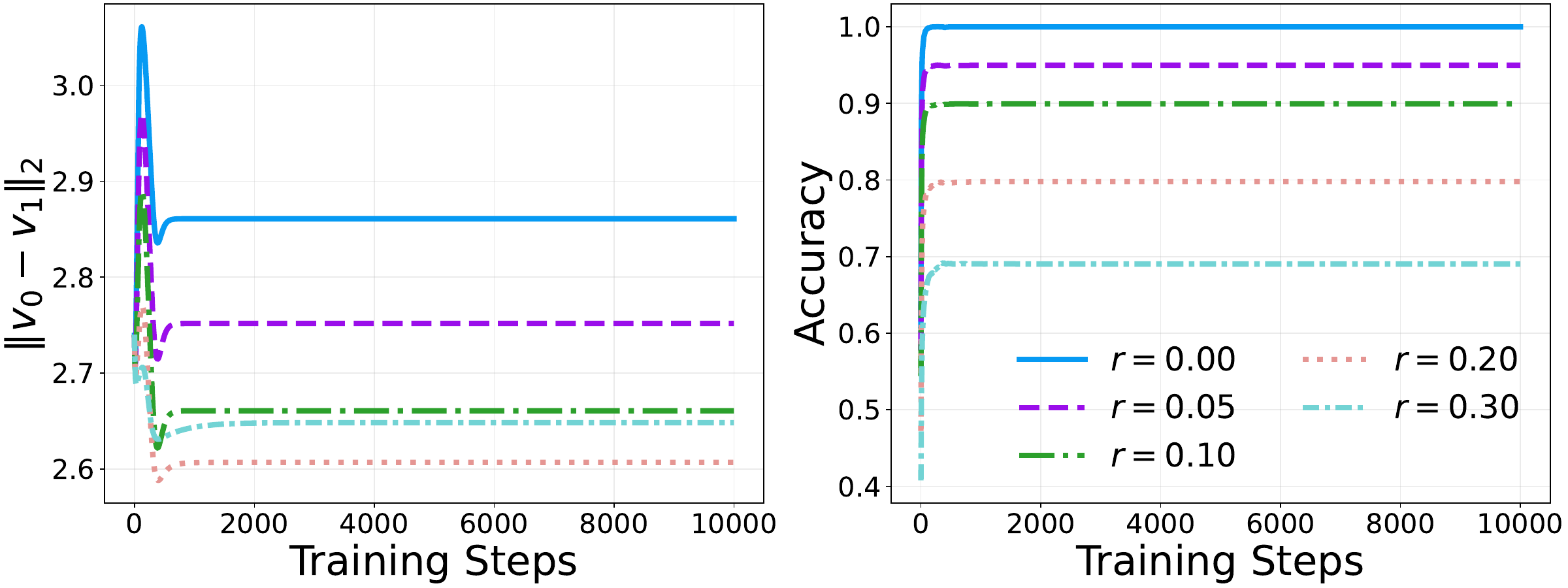}
    \caption{Training dynamics of the model with projection head and stop-gradient in the linear teacher--student model for different frustration rates $r$.
    \textbf{Left:} separation between the learned label embeddings, $\lVert v_0 - v_1 \rVert_2$, versus training steps.
    \textbf{Right:} classification accuracy versus training steps.
    }
    \label{fig:simsiam teacher student}
\end{figure}

In Fig.~\ref{fig:simsiam teacher student}, we report the training dynamics in the linear teacher--student model under stop-gradient for different frustration rates $r$, using an embedding space of dimension $5$ as in the collapse-dynamics setting.
The left panel shows the distance between the two label embeddings, $\|v_0-v_1\|_2$, which quickly reaches a plateau for all $r$, indicating that the dynamics with stop-gradient avoid complete collapse even as frustration increases.
Since we consider two classes, this separation is simply twice the distance of each label embedding from their mean.
The right panel reports the classification accuracy, which approaches the ideal value $1-r$, consistent with only the frustrated fraction of samples being intrinsically ambiguous while the remaining $(1-r)$ fraction is, in principle, perfectly classifiable.
In particular, the non-collapsing behavior in the left panel does not come at the expense of performance: near-ideal accuracy is preserved while the label separation remains nonzero.

Notably, the stop-gradient also breaks the scale equivariance of $\ell$ by removing the reciprocal coupling in the updates: because only one branch receives gradients, the scale-transformation direction that would otherwise shrink the absolute distances between embeddings, even when the geometric structure is preserved, is no longer available. As a result, the label embeddings remain at a finite separation instead of slowly approaching each other through pure rescaling, as clearly seen for $r=0$.


\section{Summary and outlook}
\label{sec:conclusion}

In this paper, we studied the widely observed phenomenon of representation collapse, where learned representations lose discriminative structure and different inputs become indistinguishable.
We analyzed this mechanism in a classification-representation setting, where sample embeddings were trained to match learned label embeddings under an alignment objective.
This setting let us quantify collapse more directly by tracking the shrinkage of pairwise separations between label embeddings.
To understand the underlying mechanism, we introduced a minimal model that optimized only the embeddings, so the dynamics and fixed points could be analyzed in closed form.
We showed both analytically and empirically that, in the perfectly classifiable case, this minimal model did not exhibit collapse.
Collapse was instead induced by \emph{frustration}, i.e. samples that could not be distinguished perfectly by the network.
Frustration introduces an additional time scale on top of the original fitting time scale. When the fraction of indistinguishable samples is small, this new time scale is slow and well separated from the original one. This separation captures the characteristic behavior observed in practice: performance improves rapidly at early times, then degrades at late times as the system drifts toward a collapsed state. Notably, the resulting two-time-scale structure in our toy model echoes the two time scales reported for diffusion models \cite{bonnaire2025why}. Here, the fast time scale corresponds to a generalization regime, where the model finds a “middle ground” that fits the data well, while the slow time scale corresponds to a memorization regime, where the model attempts to overfit the unclassifiable samples.

Under the same minimal model framework, we studied how commonly used intrinsic mechanisms prevent collapse.
By adding a shared projection head and applying stop-gradient, we showed that the dynamics admit non-collapsed fixed points that are absent without stop-gradient, thereby stabilizing finite class separation.
While the resulting dynamics were no longer exactly solvable due to the nonlinearity introduced by the projection head, we derived a set of self-consistency equations in the spirit of dynamical mean-field theory (DMFT) to characterize the evolution.
This formalism consistently explained how frustration-induced collapse was mitigated, as reflected by the fixed-point structure and supported by empirical simulations.
Finally, we demonstrated empirically that the same qualitative phenomena and collapse-prevention effects appeared in a linear teacher-student framework, indicating that the minimal model captured robust features of collapse dynamics beyond the pure embedding setting.

While our model effectively described the time-scale separation in representation collapse and how projection head and stop-gradient prevent collapse, it still missed behaviors and mechanisms that are worth exploring in future work.
In our minimal model, sample embeddings moved freely and could converge exactly to their corresponding label embeddings.
This differed from real data, where even in the best case samples from the same class form a finite-width cluster around the label embedding rather than collapsing to a single point.
As a result, our setting was not well suited for studying accuracy directly.
Once the classifiable samples collapsed to single points, accuracy remained essentially unchanged unless the \emph{label} embeddings themselves nearly coincided.
Even if their separation decayed exponentially, it did not reach zero in finite time, so any accuracy change tied to exact collapse would be effectively unobservable on practical training horizons.
A natural extension is to introduce an effective repulsive interaction among samples within the same class, which would stabilize finite clusters around each label embedding and enable a more faithful link between collapse dynamics and classification performance.

Second, we ignored several sources of stochasticity and additional couplings that are standard in modern training.
Practical optimization uses SGD rather than full-batch gradient descent, so a single run can be driven by fluctuations across basins toward a collapsed fixed point even when the \emph{average} dynamics would avoid collapse, which becomes especially relevant at large scale where retraining is costly.
Given an explicit dynamical equation together with a noise model, these stochastic effects could in principle be analyzed using the Martin--Siggia--Rose--Janssen--De Dominicis (MSRJD) formalism \cite{martin1973statistical, de1976technics, janssen1976lagrangean,tauber2014critical}.
We also did not study weight decay, momentum, or explicit manifold constraints, all of which introduce additional interactions and time scales that can qualitatively change the long-time behavior and potentially mask the effective mechanisms isolated by our minimal model.

\section*{AI Usage}
We used large language models as writing and coding assistants to improve clarity and productivity.
For writing, AI tools were used to polish grammar, refine phrasing, and suggest edits to the presentation.
For coding, AI tools were used to help draft and refactor utility functions (e.g., plotting, data loading, and experiment scaffolding) and to suggest debugging checks.
All technical content, including the modeling assumptions, theoretical derivations, experimental design, code verification, figures, and final scientific claims, was developed and validated by the authors.
The authors reviewed, edited, and take full responsibility for the final manuscript and the released code.

\begin{acknowledgments}
LHY is indebted to Tianyu Jiang for fruitful discussions and providing computational resources.
\end{acknowledgments}

\bibliographystyle{apsrev4-2}
\bibliography{ref}

\newpage
\appendix
\section{Dynamics of the Unfrustrated MSE Model}
\label{app:mse-solution}

In this appendix, we provide the detailed derivation of the gradient-flow dynamics and the fixed-point structure for the unfrustrated mean-squared-error (MSE) model introduced in Sec.~III.

We consider the loss
\begin{equation}
\ell(\{u\},\{v\})
=\frac{1}{2}\sum_{i=1}^n\sum_{\alpha=1}^N
\big(u_{i\alpha}-v_i\big)^2,
\end{equation}
where $u_{i\alpha}\in\mathbb{R}^d$ denotes the embedding of sample $\alpha$ from class $i$ and $v_i\in\mathbb{R}^d$ denotes the embedding of the corresponding class label.
Since different embedding coordinates do not couple, the dynamics can be analyzed component-wise.
We therefore restrict to the one-dimensional case without loss of generality.

Under full-batch gradient descent with learning rate $\tilde{\gamma}$, the discrete-time updates are
\begin{equation}
\begin{aligned}
\Delta u_{i\alpha}
&=-\tilde{\gamma}\big(u_{i\alpha}-v_i\big),\\
\Delta v_i
&=-\tilde{\gamma}\sum_{\alpha=1}^N
\big(v_i-u_{i\alpha}\big).
\end{aligned}
\end{equation}
Taking the continuous-time limit with normalized learning rate $\gamma$, we obtain the gradient-flow equations
\begin{equation}
\begin{aligned}
\dot{u}_{i\alpha}
&=-\gamma\big(u_{i\alpha}-v_i\big),\\
\dot{v}_i
&=-\gamma\sum_{\alpha=1}^N
\big(v_i-u_{i\alpha}\big).
\end{aligned}
\label{eq:mse-flow-app}
\end{equation}

A key feature of Eq.~\eqref{eq:mse-flow-app} is that the dynamics are fully decoupled across classes.
All forces acting on $(u_{i\alpha},v_i)$ depend only on the class index $i$, and there is no coupling between different classes.
We therefore analyze the dynamics for a fixed class $i$ and suppress the class index where no confusion arises.

Define the within-class sum
\begin{equation}
U \equiv \sum_{\alpha=1}^N u_{\alpha},
\end{equation}
so that summing Eq.~\eqref{eq:mse-flow-app} over $\alpha$ yields
\begin{equation}
\begin{aligned}
\dot{U}
&=-\gamma\big(U-N v\big),\\
\dot{v}
&=-\gamma\big(N v-U\big).
\end{aligned}
\label{eq:Uv-flow}
\end{equation}
The fixed-point manifold of this system is given by
\begin{equation}
U=N v,
\end{equation}
corresponding to perfect alignment between sample embeddings and the label embedding.

To analyze relaxation toward this manifold, define the deviation
\begin{equation}
\delta(t)\equiv U(t)-N v(t).
\end{equation}
Using Eq.~\eqref{eq:Uv-flow}, its time derivative is
\begin{equation}
\begin{aligned}
\dot{\delta}
&=\dot{U}-N\dot{v}\\
&=-\gamma\big(U-N v\big)
+\gamma N\big(N v-U\big)\\
&=-\gamma(N+1) \delta.
\end{aligned}
\end{equation}
Thus,
\begin{equation}
\delta(t)=\delta(0) 
\mathrm{e}^{-\gamma(N+1)t},
\end{equation}
where $\delta(0)=U(0)-N v(0)$.

Since $\dot{v}=\gamma\delta$, integration gives
\begin{equation}
\begin{aligned}
v(t)
&=v(0)
+\gamma\int_0^t \delta(s) ds\\
&=v(0)
+\frac{\delta(0)}{N+1}
\Big(1-\mathrm{e}^{-\gamma(N+1)t}\Big).
\end{aligned}
\vspace{1mm}
\end{equation}
Taking the long-time limit $t\to\infty$, we obtain
\begin{equation}
v(\infty)
=v(0)
+\frac{U(0)-N v(0)}{N+1}
=\frac{U(0)+v(0)}{N+1}.
\end{equation}
Writing $U(0)=N\bar{u}(0)$ in terms of the initial within-class mean
\begin{equation}
\bar{u}(0)
=\frac{1}{N}\sum_{\alpha=1}^N u_{\alpha}(0),
\end{equation}
yields the fixed-point value quoted in the main text,
\begin{equation}
v(\infty)
=\frac{N \bar{u}(0)+v(0)}{N+1}.
\end{equation}

Finally, substituting $v(\infty)$ back into the solution for $u_{\alpha}$ shows that
\begin{equation}
u_{\alpha}(t)\;\longrightarrow\;v(\infty)
\qquad
\text{for all }\alpha,
\end{equation}
demonstrating that all sample embeddings within a class align with the label embedding, while different classes remain independent.
This completes the derivation of the unfrustrated dynamics.


\section{Exact Spectrum of the Frustrated Dynamics}
\label{sec:full-spectrum}

In this section we derive the full mode decomposition and eigenvalues of
Eq.~\eqref{eq:frustrated-dynamics-noproj}.
Although the system has $\mathcal{O}(nN)$ degrees of freedom, its permutation
symmetries imply an exact block-diagonalization into invariant subspaces.
Throughout we assume $r>0$ so that the shared-sample variables are well-defined.

\subsection{Decomposition of \texorpdfstring{$u_{i\alpha}$ and $s_\alpha$}{uialpha and salpha} into Means and Fluctuations}

Define the within-class mean and fluctuations of the non-shared samples:
\begin{equation}
\bar u_i
\equiv
\frac{1}{(1-r)N}\sum_{\alpha=1}^{(1-r)N} u_{i\alpha},
\qquad
\delta u_{i\alpha}
\equiv
u_{i\alpha}-\bar u_i,
\end{equation}
so that $\sum_{\alpha}\delta u_{i\alpha}=0$ for each $i$.
Similarly, define the shared-sample mean and fluctuations:
\begin{equation}
\bar s
\equiv
\frac{1}{rN}\sum_{\alpha=1}^{rN} s_{\alpha},
\qquad
\delta s_{\alpha}
\equiv
s_{\alpha}-\bar s,
\end{equation}
so that $\sum_{\alpha}\delta s_{\alpha}=0$.

Averaging Eq.~\eqref{eq:frustrated-dynamics-noproj} over $\alpha$ yields the
closed dynamics for the means:
\begin{align}
\dot{\bar u}_i
&=
-\gamma(\bar u_i-v_i),
\label{eq:ubar_mean}\\
\dot{\bar s}
&=
-\gamma\Big(n\bar s-\sum_{j=1}^n v_j\Big),
\label{eq:sbar_mean}\\
\dot v_i
&=
-\gamma\Big(
Nv_i-(1-r)N \bar u_i-rN \bar s
\Big).
\label{eq:vi_mean}
\end{align}

Subtracting \eqref{eq:ubar_mean} from the equation for $u_{i\alpha}$ gives
the fluctuation dynamics
\begin{equation}
\dot{\delta u}_{i\alpha}
=
-\gamma \delta u_{i\alpha}.
\label{eq:utilde}
\end{equation}
Likewise, subtracting \eqref{eq:sbar_mean} from the equation for $s_\alpha$ gives
\begin{equation}
\dot{\delta s}_{\alpha}
=
-\gamma n \delta s_{\alpha}.
\label{eq:stilde}
\end{equation}
Equations \eqref{eq:utilde}--\eqref{eq:stilde} show that the internal sample
fluctuations decouple exactly from $(\bar u_i,\bar s,v_i)$.

\subsection{Eigenvalues of the Fluctuation Subspaces}

Equation \eqref{eq:utilde} implies that each independent component of
$\delta u_{i\alpha}$ decays with eigenvalue
\begin{equation}
\lambda_u = -\gamma.
\end{equation}
The multiplicity is $n\big((1-r)N-1\big)$ (one constraint $\sum_\alpha \delta u_{i\alpha}=0$
per class).

Equation \eqref{eq:stilde} implies that each independent component of
$\delta s_{\alpha}$ decays with eigenvalue
\begin{equation}
\lambda_s = -\gamma n,
\end{equation}
with multiplicity $(rN-1)$ (one constraint $\sum_\alpha \delta s_\alpha=0$).

These modes describe within-group dispersion and do not affect class contrast.

\subsection{Reduction of the Mean Dynamics into Deviation and Mean Sectors}

The remaining dynamics \eqref{eq:ubar_mean}--\eqref{eq:vi_mean} live in the
$(2n+1)$-dimensional space of $(\bar u_i,v_i,\bar s)$.
Define global means and deviations across classes:
\begin{equation}
\bar{u} \equiv \frac{1}{n}\sum_{i=1}^n \bar u_i,
\qquad
\bar{v} \equiv \frac{1}{n}\sum_{i=1}^n v_i,
\end{equation}
\begin{equation}
\delta u_i \equiv \bar u_i-\bar{u},
\qquad
\delta v_i \equiv v_i-\bar{v},
\qquad
\sum_{i=1}^n \delta u_i=\sum_{i=1}^n \delta v_i=0.
\end{equation}
Substituting into \eqref{eq:ubar_mean}--\eqref{eq:vi_mean} yields two independent
blocks:

\paragraph{(i) Deviation sector.}
For each of the $(n-1)$ independent deviation directions,
\begin{equation}
\frac{d}{dt}
\begin{pmatrix}
\delta u\\
\delta v
\end{pmatrix}
=
\gamma
\begin{pmatrix}
-1 & 1\\
N(1-r) & -N
\end{pmatrix}
\begin{pmatrix}
\delta u\\
\delta v
\end{pmatrix},
\label{eq:dev_block}
\end{equation}
with eigenvalues
\begin{equation}
\lambda^{\mathrm{dev}}_{\pm}
=
-\frac{\gamma}{2}
\Big[
N+1
\pm
\sqrt{(N+1)^2-4Nr}
\Big].
\label{eq:dev_eigs_full}
\end{equation}
Each of $\lambda^{\mathrm{dev}}_{\pm}$ has multiplicity $(n-1)$.
These modes control the decay of inter-class contrasts, e.g.\ $v_i-\bar{v}$.

\paragraph{(ii) Mean sector.}
The triplet $(\bar{u},\bar{v},\bar s)$ evolves as
\begin{equation}
\frac{d}{dt}
\begin{pmatrix}
\bar{u}\\
\bar{v}\\
\bar s
\end{pmatrix}
=
\gamma
\begin{pmatrix}
-1 & 1 & 0\\
N(1-r) & -N & Nr\\
0 & n & -n
\end{pmatrix}
\begin{pmatrix}
\bar{u}\\
\bar{v}\\
\bar s
\end{pmatrix}.
\label{eq:mean_block}
\end{equation}
This block has one exact zero eigenvalue,
\begin{equation}
\lambda^{\mathrm{mean}}_0=0,
\end{equation}
reflecting the global translation symmetry of the squared loss.
A convenient conserved quantity is
\begin{equation}
r \bar s(t)
+
\frac{n}{N} \bar{v}(t)
+
n(1-r) \bar{u}(t)
=
\mathrm{const}.
\label{eq:mean_conserved_full}
\end{equation}
The remaining two eigenvalues are
\begin{equation}
\lambda^{\mathrm{mean}}_{\pm}
=
-\frac{\gamma}{2}
\Big[
N+n+1
\pm
\sqrt{(N-n+1)^2+4Nr(n-1)}
\Big].
\label{eq:mean_eigs_full}
\end{equation}

\subsection{Full Spectrum and Multiplicities}

Collecting the invariant subspaces, the spectrum of the full linear system
\eqref{eq:frustrated-dynamics-noproj} consists of:
\begin{itemize}
\item $\lambda_u=-\gamma$ with multiplicity
$n\big((1-r)N-1\big)$ \hfill (within-class $u$ fluctuations);
\item $\lambda_s=-\gamma n$ with multiplicity
$(rN-1)$ \hfill (within-shared $s$ fluctuations);
\item $\lambda^{\mathrm{dev}}_{\pm}$ from \eqref{eq:dev_eigs_full},
each with multiplicity $(n-1)$ \hfill (inter-class deviation modes);
\item $\lambda^{\mathrm{mean}}_{\pm}$ from \eqref{eq:mean_eigs_full},
each with multiplicity $1$, and a neutral mode $\lambda^{\mathrm{mean}}_0=0$
\hfill (global mean sector).
\end{itemize}

In particular, the inter-class contrasts $v_i-\bar{v}$ lie entirely in the deviation
sector and are governed only by $\lambda^{\mathrm{dev}}_{\pm}$.
The alignment errors $\bar u_i-v_i$ receive contributions from both the deviation
and mean sectors, while internal sample dispersions decay independently according
to $\lambda_u$ (and $\lambda_s$ for shared-sample dispersion).

\end{document}